\documentclass[twocolumn]{aastex62}
\pdfoutput=1

\newcommand\Gaia{{\it Gaia }}

\usepackage{amsmath,graphicx,multirow}
\usepackage{ulem}
 
\received{January 24, 2019}
\revised{May 15, 2019}
\accepted{May 15, 2019}
\submitjournal{ApJ}

\shorttitle{Short Term Variability of Evolved Massive Stars with TESS}
\shortauthors{Dorn-Wallenstein, Levesque, \& Davenport}

\begin{document}

\title{Short Term Variability of Evolved Massive Stars with TESS}

\correspondingauthor{Trevor Z. Dorn-Wallenstein}
\email{tzdw@uw.edu}

\author[0000-0003-3601-3180]{Trevor Z. Dorn-Wallenstein}
\affiliation{University of Washington Astronomy Department \\
Physics and Astronomy Building, 3910 15th Ave NE  \\
Seattle, WA 98105, USA} 

\author[0000-0003-2184-1581]{Emily M. Levesque}
\affiliation{University of Washington Astronomy Department \\
Physics and Astronomy Building, 3910 15th Ave NE  \\
Seattle, WA 98105, USA}

\author[0000-0002-0637-835X]{James R. A. Davenport}
\altaffiliation{DIRAC Fellow}
\affiliation{University of Washington Astronomy Department \\
Physics and Astronomy Building, 3910 15th Ave NE  \\
Seattle9, WA 98105, USA}

\begin{abstract}

We present the first results from a study of TESS Sector 1 and 2 light curves for eight evolved massive stars in the LMC: six yellow supergiants (YSGs) and two luminous blue variables (LBVs), including S Doradus. We use an iterative prewhitening procedure to characterize the short-timescale variability in all eight stars. The periodogram of one of the YSGs, HD 269953, displays multiple strong peaks at higher frequencies than its fellows. While the field surrounding HD 269953 is quite crowded, it is the brightest star in the region, and has infrared colors indicating it is dusty. We suggest HD 269953 may be in a post-red supergiant evolutionary phase. We find a signal with a period of $\sim5$ days for the LBV HD 269582. The periodogram of S Doradus shows a complicated structure, with peaks below frequencies of 1.5 cycles per day. We fit the shape of the background noise of all eight light curves, and find a red noise component in all of them. However, the power law slope of the red noise and the timescale over which coherent structures arise changes from star to star. Our results highlight the potential for studying evolved massive stars with TESS.

\end{abstract}


\section{Introduction} \label{sec:intro}

The environments in and around evolved massive stars are complex and unique astrophysical laboratories. Much of the information about the physics of these stars is encoded within their variability. However, due to their rarity, the behavior of massive stars in the time domain is still poorly studied by high-precision space-based instruments. Thus, the critical physical ingredients that inform our models of evolved massive stars (e.g., the distribution of rotation rates, asteroseismically determined masses and radii, short-timescale wind-driven variability and more) are still poorly constrained by observations.

On the main sequence, massive stars manifest themselves as O and B dwarfs earlier than spectral type $\sim$B3 \citep{habets81}. During and shortly after the main sequence phase (i.e., OB dwarfs and supergiants), mass loss rates are at their lowest \citep{puls08,smith14}, and the geometries of their circumstellar media (CSM) are at their simplest \citep[e.g., ][]{garciasegura96,gvaramadze18}. Thus, rotational modulation from surface features \citep[e.g.,][]{aerts13,ramiaramanantsoa18} or Co-rotating Interaction Regions \citep[CIRs, see ][]{mullan84,cranmer96} in the stellar wind can be readily observed by {\it CoRoT}, {\it Kepler}, and K2 \citep[see ][ and more]{blomme11,buysschaert15,balona15,balona16,johnston17}, and has been observed in the B supergiant HD 46769 \citep{aerts13}.

At shorter timescales, oscillations in main sequence B (and more recently O) stars have been detected from space \citep[e.g., ][]{balona11,blomme11,buysschaert15,johnston17} as p-modes in $\beta$ Cephei pulsators, g-modes in Slowly Pulsating B-type (SPB) stars, or a combination of both \citep{daszynska18}. There are also sources of stochastic or noncoherent variability \citep{blomme11} that could arise due to the sub-surface convection zone (which may interact with stellar pulsations, see \citealt{perdang09}), granulation, or inhomogeneities in the stellar wind. Additional variability may manifest itself at $\sim$hour timescales due to instabilities in the stellar wind \citep{kritcka18}. Finally, pulsations in massive stars OB supergiants have been studied both observationally \citep{saio06,aerts17} and theoretically \citep[e.g., ][]{godart09,daszynska13,ostrowski15,ostrowski17}.

Beyond these early phases, massive stars are poorly understood at short timescales. \citet{conroy18} demonstrated that these stars display a rich variety of variability on day to decade timescales. However, no massive stars evolved beyond the blue supergiant phase were observed at higher cadence by {\it Kepler} or K2, and only small samples of evolved stars at specific evolutionary phases were have been observed with targeted campaigns using {\it CoRoT} or the BRITE constellation. That said, these stars are fantastic targets for high cadence photometry. In red supergiants (RSGs), convective and pulsational processes can generate variability on long timescales \citep{wood83}, which helps RSGs launch dusty stellar winds \citep[and references therein]{yoon10}. Simulations of red supergiants \citep{chiavassa11} predict large scale convective motions, turbulence, and shocks, all of which can manifest themselves coherently or stochastically (e.g., red noise detected in AAVSO light curves of RSGs by, \citealt{kiss06}, or the longer pulsations predicted by \citealt{yoon10}).

Studies of photometric variability in Wolf-Rayet stars are still relatively few in number. The BRITE constellation has studied six of the brightest Wolf-Rayet stars \citep{moffat18}, and detected CIRs, binary interactions, and stochastic variability. However, with such a small sample size, little can be said about how the variability of Wolf-Rayets depends on fundamental stellar parameters like temperature and luminosity. Finally, stars in transitional states with lifetimes of only a few $10^4$ years (e.g., Yellow Supergiants, luminous blue variables, ``slash'' stars, etc.) have gone completely unobserved, due to their rarity and thus their lack of concentration on the sky; any pointing by a mission with a stationary field of view (e.g., {\it Kepler}) isn't likely to include many short-lived evolutionary phases of massive stars. However, many of these evolutionary phases are still poorly understood. Many of them are associated with dusty circumstellar mediums, outbursts, and other phenomena. Their pulsational or rotational properties can be used to infer information about their interior states and evolution, including angular momentum transport, convection, surface differential rotation and more. 

The Transiting Exoplanet Survey Satellite (TESS, \citealt{ricker15}) is a nearly all-sky photometry mission targeting $\sim$20,000 bright stars per year at a two-minute cadence (with full-frame images for $\sim$20 million stars every thirty minutes), yielding approximately 27 days of continuous photometry for stars close to the ecliptic plane, with longer light curves for stars observed by multiple spacecraft pointings. Large numbers of evolved massive stars in the Galaxy and the Magellanic Clouds are bright enough to be observed by TESS over the course of its nominal two-year mission. Here we present analysis of the first evolved massive star light curves to become available from TESS sectors 1 and 2. In \S\ref{sec:sample}, we discuss our sample selection using data from the \Gaia mission. Results for each star are presented in \S\ref{sec:results}. We discuss the relevance of our findings for stellar evolution theory, and the prospects of a dedicated TESS campaign to observe evolved massive stars in \S\ref{sec:discuss}, before concluding in \S\ref{sec:conclusion}.

\section{Sample Selection and Data Processing}\label{sec:sample}

We first attempted to search for evolved massive stars in our Galaxy, using the accurate astrometry published in the second data release (DR2) of the \Gaia \citep{gaia18}. DR2 contains position and brightness measurements in the broad \Gaia passband $G$ for 1.7 billion stars, of which 1.4 billion have photometry in the blue and red bandpasses $G_{BP}$ and $G_{RP}$, and 1.3 billion have parallax $\varpi$ and proper motion $\mu$ measurements. We acquired the TESS Sectors 1 and 2 target lists\footnote{Target lists obtainable at https://tess.mit.edu/observations/sector-1/ and https://tess.mit.edu/observations/sector-2/ respectively}, uploaded them to the ESA \Gaia archive\footnote{https://gea.esac.esa.int/archive/}, and searched for objects in \Gaia DR2 and the TESS target lists that were separated by less than 1''. 

In theory, \Gaia parallaxes are easily convertible to distances via
\begin{equation}
    \frac{d}{\rm pc} = \frac{\rm arcsec}{\varpi}
\end{equation}
which would allow for a direct measurement of luminosity, and then used to select massive stars. However, converting from parallax to distance is a nontrivial task in practice. Systematics --- e.g., parallax and proper motion zero-point offsets measured from distant QSOs --- exist in the data \citep{lindegren18}, and many objects have high fractional errors ($\sigma_\varpi/\varpi$) or negative measured parallax. In a Bayesian framework these measurements are all useful, and \citet{bailerjones18} inferred distances for the majority of stars in \Gaia DR2 accounting for these effects, using a prior based on the spatial distribution of stars in the galaxy. For stars in the TESS--\Gaia cross-match, we calculated the absolute $G$ magnitude:
\begin{equation}
M_G = G - 5\log_{10}r_{est} + 5 - A_G    
\end{equation}
using the estimated distance $r_{est}$ from \citet{bailerjones18}, and the published estimate of the extinction $A_G$. We also estimated the reddening as
\begin{equation}
E(G_{BP} - G_{RP}) = \frac{A_G}{A_G/A_V}\Big(\frac{A_{BP}}{A_V} - \frac{A_{RP}}{A_V}\Big) 
\end{equation}
using the estimated extinction $A_G$, and coefficients from \citet{malhan18}. We make the cross-matched data, as well as the estimated $M_G$ and $G_{BP}-G_{RP}$ publicly available online.\footnote{https://github.com/tzdwi/TESS-Gaia}For stars without an estimated $A_G$, we also estimate a lower limit to the absolute magnitude
\begin{equation}
    M_G \leq G - 5\log_{10}r_{est} + 5
\end{equation}
and an upper limit to the intrinsic $G_{BP}-G_{RP}$ by assuming $E(G_{BP}-G_{RP}) = 0$. 

\begin{deluxetable*}{lcrllllll}
\tabletypesize{\footnotesize}
\tablecaption{TESS two-minute cadence targets. TIC \# and T magnitude are from the TIC; J  magnitude is from 2MASS \citep{cutri03}, and 3.6, 4.5, 5.8, and 8.0 $\mu$m magnitudes are from the Spitzer SAGE LMC survey \citep{bonanos09}\label{tab:sample}.}
\tablehead{\colhead{Common Name} & \colhead{Evolutionary Stage} & \colhead{TIC \#} & \colhead{$T$} & \colhead{$J$} &  \colhead{$J - [3.6]$}& \colhead{$J - [4.5]$} & \colhead{$J - [5.8]$} & \colhead{$J - [8.0]$} \\
\colhead{} & \colhead{} & \colhead{} & \colhead{[mag]} & \colhead{[mag]} & \colhead{[mag]} & \colhead{[mag]} & \colhead{[mag]} & \colhead{[mag]}} 
\startdata
S Dor & LBV & 179305185 & 9.16 & 8.683  & 0.923 & 1.036 & 1.164  & 1.373 \\
HD 269953 & YSG & 404850274 & 9.22 & 8.588 &  1.311 & 1.89 & 2.363  & 3.427 \\
HD 269582 & Ofpe/WN9 $\rightarrow$ LBV & 279957111 & 9.33 & 12.041 & 2.814 & 3.188 & 3.255 & 3.612 \\
HD 270046 & YSG & 389437365 & 9.45 & 8.713 & 0.712 & 0.748 & 0.933  & 0.894 \\
HD 270111 & YSG & 389565293 & 9.63 & 9.073 & 0.535 & 0.578 & 0.599 & 0.480 \\
HD 269331 & YSG & 179206253 & 9.82 & 9.594 & 0.373 & 0.457 & 0.525 & 0.580 \\
HD 269110 & YSG & 40404470 & 10.01 & 9.320 & 0.560 & 0.643 & 0.695 & 0.891 \\
HD 268687 & YSG & 29984014 & 10.21 & 9.693 & 0.397 & 0.523 & 0.608  & 0.706 \\
\enddata
\end{deluxetable*}

Because Sectors 1 and 2 contain both the Small and Large Magellanic Clouds (MCs), the Galactic prior used by \citet{bailerjones18} is inappropriate for a subset of our sample. \citet{gaiacollab18} used data from Gaia DR2 to select likely MC members. For these stars, we adopt distance moduli of 19.05/18.52 for the SMC/LMC respectively \citep{kovacs00a,kovacs00}, and assume the values of $R_V$ from \citet{gordon03} and $E(B-V)$ from \citet{massey07} to calculate the average $A_G$ and $E(G_{BP}-G_{RP})$ towards both MCs.

\begin{figure}[ht!]
\includegraphics[width=3.7in]{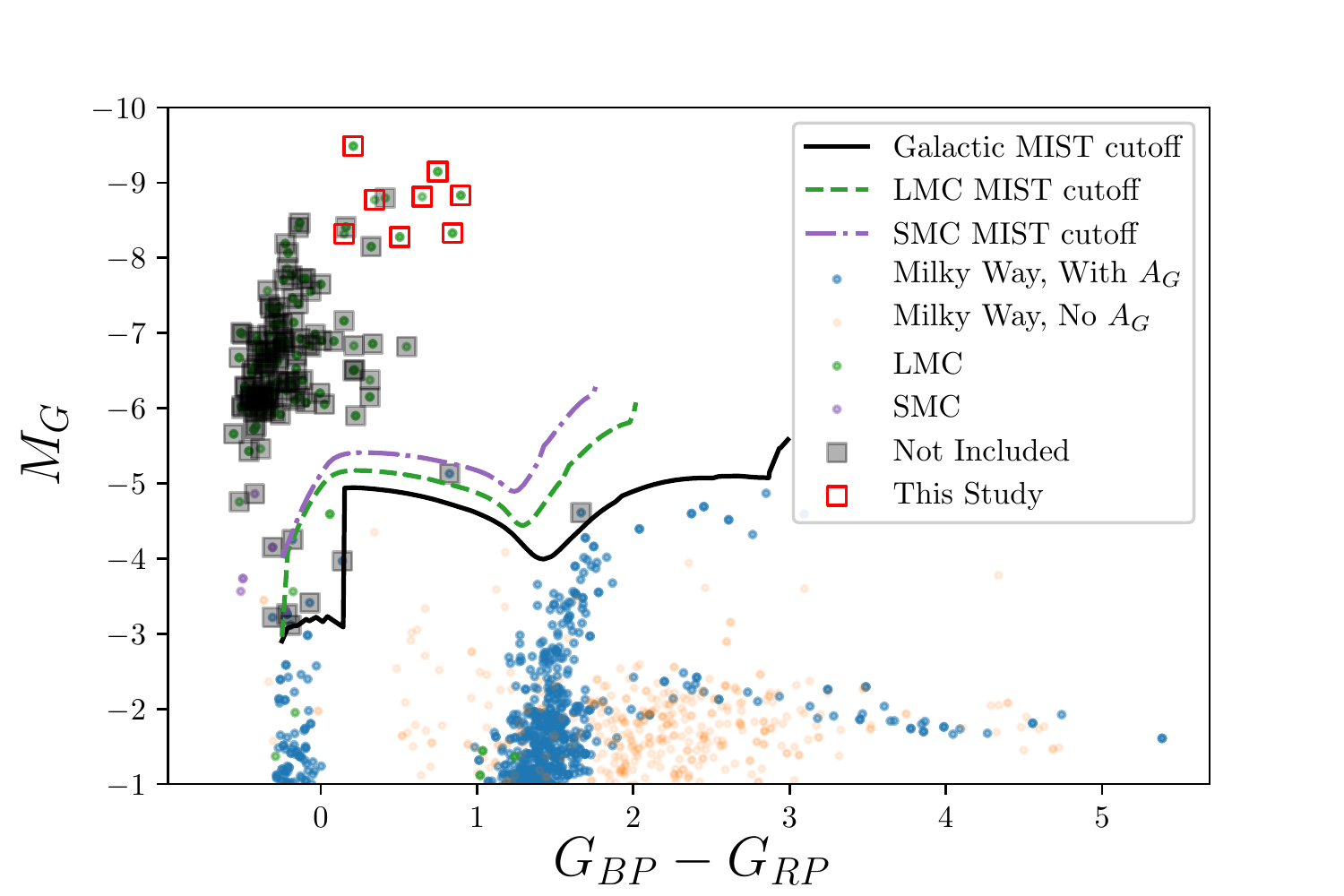}
\caption{\Gaia CMD for TESS Sector 1 and 2 targets. Galactic stars with an estimate of $A_G$ are in blue, while stars without an $A_G$ estimate are in orange; for these stars, colors are upper limits, and magnitudes are lower limits. Points in green (purple) are in the LMC (SMC) The black, green, and purple lines represent our minimum-luminosity criteria to select massive stars in the Galaxy, LMC, and SMC respectively. Stars in grey boxes are either low mass stars, or relatively unevolved O and B stars. The red boxes indicate the eight evolved massive stars we select for this study.}\label{fig:cmd}
\end{figure}

We can then construct accurate color-magnitude diagrams (CMDs), which we can use to select massive stars from targets observed by TESS. We use isochrones from the MESA Isochrones \& Stellar Tracks (MIST, \citealt{dotter16,choi16}) group, which adopts stellar models from the Modules for Experiments in Stellar Astrophysics (MESA, \citealt{paxton11,paxton13,paxton15}) code. In particular we chose isochrones with metallicity $[Fe/H] = 0.25$ for the Galaxy, and rotation speed relative to critical of $v/v_{crit}=0.4$ --- the $[Fe/H] = 0.25$ isochrones were chosen to follow the general distribution of main sequence stars, which we wish to avoid in our sample \citep[e.g.,][]{davenport2018}. For the LMC (SMC) we choose the corresponding $[Fe/H] = -0.5$ (-1) isochrones. We then selected the faintest isochrone point of any age between $10^5$ and $10^{10.3}$ yr in steps of 0.05 dex with initial mass $M_i \geq 8$ M$_\odot$ in small bins of $G_{BP}-G_{RP}$. This essentially forms a boundary in color-magnitude space that represents the faintest luminosities reached by any massive star at any point during its evolution, and no fainter massive stars are expected to be found --- note that many isochrone points with $M_i < 8$ M$_\odot$ lie above this boundary, so our sample is not constructed to be free of contamination. We show the boundary for each metallicity isochrone set, as well as \Gaia colors and absolute magnitudes in Figure \ref{fig:cmd}. Points in blue are stars for which our estimate of $M_G$ and $G_{BP}-G_{RP}$ include the extinction, and stars in orange are those without estimates of $A_G$ in \Gaia DR2. Stars in green (purple) are stars belonging to the LMC (SMC), as identified by \citet{gaiacollab18}. The black/green/purple lines denotes our luminosity cutoff for selecting massive stars in the Galaxy/LMC/SMC.

Of these stars, many are Galactic long period variables (LPVs), and a number are main sequence or giant OB stars, as well as some Be stars, and all were observed for specific OB or Be asteroseismology programs \citep{pedersen19}. We also remove several targets that are not spectroscopically confirmed as massive stars, located in extremely crowded fields (with multiple bright targets located in a single 21’' TESS pixel), or that are members of multiple systems (which will be studied in future work).

This leaves us with a total of eight evolved massive stars in Sectors 1 and 2 observed at two-minute cadence with TESS. All targets belong to the LMC.\footnote{As an aside: the fact that all of these stars are in the LMC emphasizes the need for TESS programs targeting Galactic evolved massive stars, where the targets are brighter and less crowded.} We list the evolutionary stage, $T$ magnitude and ID number from the TESS Input Catalog \citep[TIC, ][]{stassun18}, 2MASS $J$ magnitude \citep{cutri03}, and 2MASS/IRAC colors using data from \citet{bonanos09} in Table \ref{tab:sample}. We indicate the locations of these stars with red boxes in Figure \ref{fig:cmd}. Points outlined in grey boxes are either low mass AGB LPVs or relatively unevolved O and B stars that we ignore for this study. We note that, while this CMD-based selection was redundant for selecting our sample of luminous and well-known LMC stars, it is useful for selecting Galactic stars.

\subsection{Data Cleaning}

\begin{figure*}[ht!]
\centering
\includegraphics[width=\textwidth]{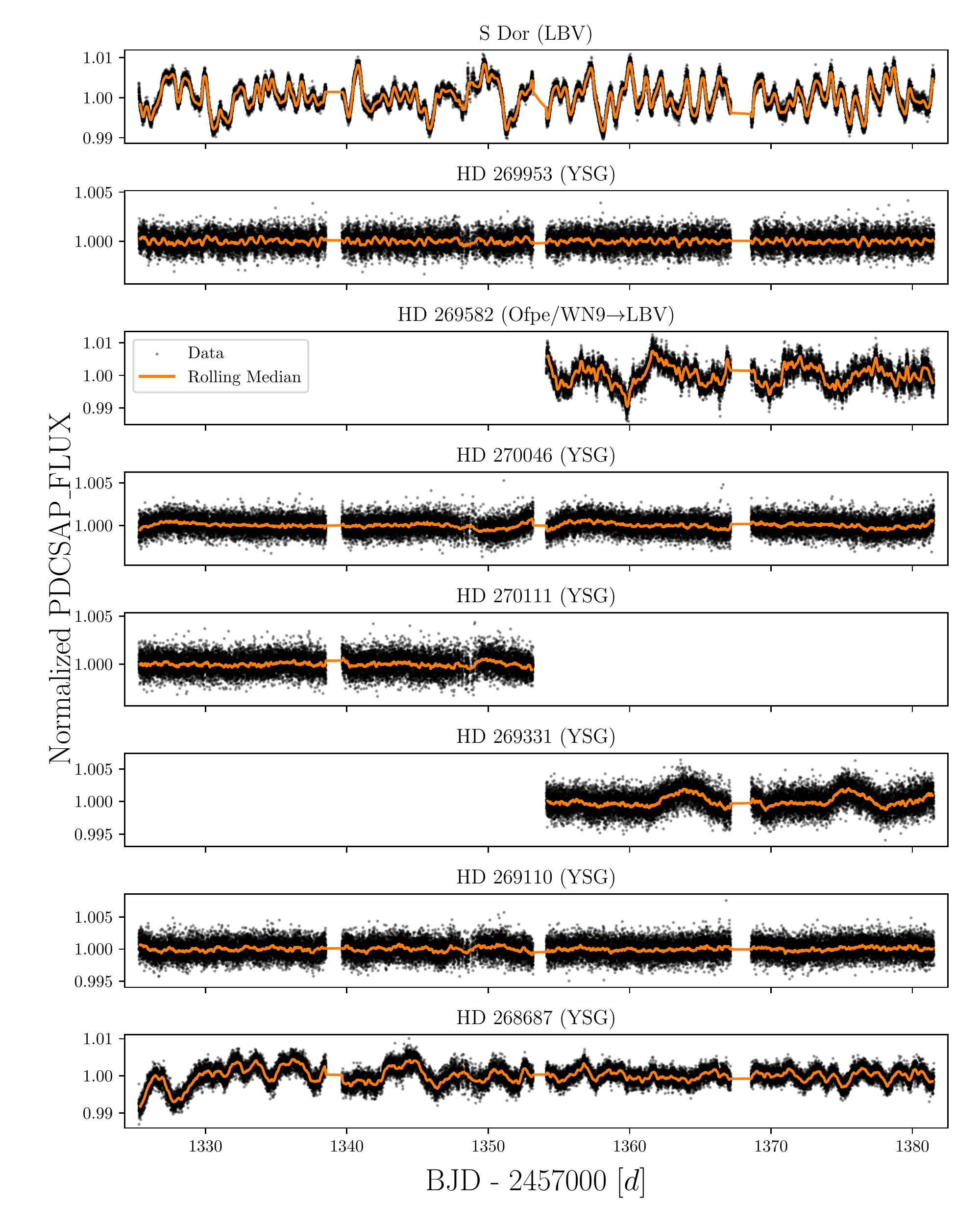}
\caption{Normalized TESS light curves for the eight target stars. Data are in black points, and a rolling 128-point median is plotted in orange.}
\label{fig:lcs}
\end{figure*}

The TESS team released raw light curves and full-frame images from Sectors 1 and 2 on 6 December 2018. We downloaded all light curves for stars observed at two-minute cadence from The Mikulski Archive for Space Telescopes (MAST), and selected the light curves associated with the TIC numbers of our targets. Because these light curves are processed by the first iteration of the TESS pipeline, we err on the side of caution, assuming that the raw light curves contain numerous instrumental effects. Thus we select the {\tt PDCSAP\_FLUX} lightcurves, which have been corrected for some systematics. We then normalize by dividing the data by the median flux. For targets observed in both Sectors 1 and 2, we choose to median-divide each Sector individually before concatenating the light curves. While this helps to eliminate Sector-to-Sector offsets and systematics, it can also erase variations at timescales longer than $\sim$1 month. We plot all of the normalized light curves, along with a rolling 128-point median in orange, in Figure \ref{fig:lcs}. Finally, for HD 268687, we fit the light curve with a 7$^{\rm th}$-order polynomial, and normalize the data by the fit in order to remove the increase in flux at the beginning of the light curve that would mask otherwise interesting behavior. This effectively acts as a low-pass filter in the Fourier domain. The resulting data have pseudo-Nyquist frequencies $f_{Ny}$ (calculated from the average time-difference between data points) between $320$ and $330$ day$^{-1}$. Due to the $\sim30$-day observing window per TESS sector, the expected width of peaks in the periodograms presented (the Rayleigh resolution, defined as the inverse of the observing baseline $T$) is $0.036$ day$^{-1}$ for HD 269582, HD 270111, and HD 269331, and $0.018$ day$^{-1}$ for the remaining stars. 

\subsection{Iterative Prewhitening}\label{subsec:prewhitening}

We now wish to describe the variability of each star in terms of sinusoids. For this, we apply the following prewhitening procedure for each lightcurve, as described in \citet{blomme11}. We first subtract the mean value of the flux. We then use Astropy \citep{astropy13,astropy18} to calculate the Lomb-Scargle Periodogram \citep{lomb76,scargle82} on the unsmoothed data for frequencies between 1/30 day$^{-1}$ and $f_{Ny}$, adopting the default Astropy grid-spacing heuristic of 5 times the Rayleigh resolution. The lower frequency limit is set to avoid overinterpreting low-frequency systematics that may exist between TESS sectors. Using the \texttt{psd} normalization, the resulting periodograms are in units of power.

At the $j^{\rm th}$ stage of prewhitening, we calculate the periodogram, and select the frequency of the peak with the highest power, $f_{max}$. We then use the \texttt{curve\_fit} routine in SciPy \citet{scipy:2001} to fit the current prewhitened flux at each time $t_i$ with a sin function $A_j \sin(2\pi f_j t_i + \phi_j)$, where $A_j$ is the semi-amplitude, $f_j$ is the frequency, and $\phi_j$ is the phase. $A_j$ and $\phi_j$ are allowed to vary freely, and $f_j$ is bound to the range $f_{max}\pm 1/T$. To calculate the errors on each parameter, we use the formulae given in \citet{lucy71} and \citet{montgomery99}:
\begin{align}
    \epsilon(f_j) &= \sqrt{\frac{6}{N}}\frac{1}{\pi T}\frac{\sigma_j}{A_j} \\
    \epsilon(A_j) &= \sqrt{\frac{2}{N}}\sigma_j \\
    \epsilon(\phi_j) &= \sqrt{\frac{2}{N}}\frac{\sigma_j}{A_j} \\
\end{align}
where $\sigma_j$ is the standard deviation of the flux at the $j^{\rm th}$ prewhitening stage.

We then subtract the fit from the flux and begin the next phase. As a stopping criterion, we adopt the Bayesian Information Content (BIC, \citealt{schwarz78})
\begin{equation}\label{eq:bic}
    BIC = -2\ln(\mathcal{L}) + m\ln(N)
\end{equation}
where $m$ is the total number of terms in the fit ($\equiv3j$), $N$ is the number of points in the lightcurve, and $\mathcal{L}$ is the likelihood, defined such that
\begin{equation}
    -2\ln(\mathcal{L}) = \sum_{i=1}^{N}\frac{(y_i - F(t_i,\Theta_m))^2}{\sigma_i^2}
\end{equation}
to within a constant, where $y_i$ are the original normalized fluxes, $F(t_i,\Theta_m)$ is the sum of all of the fit sinusoids in the current and preceding prewhitening stages evaluated at times $t_i$ and fit parameters $\Theta_m$, and $\sigma_i$ are the normalized errors in the original light curve \citep[Sect. 15.1]{press92}. To determine when we have reached the noise level of the light curve, we continue prewhitening until we reach a minimum in the fit's BIC. 

This procedure results in a list of frequencies, amplitudes, and phases that can, in principle, describe the variability of each star. However, a number of the derived frequencies are quite similar to each other (i.e., the difference in frequencies is within the Rayleigh resolution). These similar and spurious frequencies can arise due to the short length of the observing baseline \citep{loumos78}. Therefore, we filter the list of frequencies by imposing a requirement that frequencies must be separated by more than $1.5/T$. In cases where such pairs of similar frequencies are found, we discard the frequency found at the later stage of prewhitening. We list $N$, $f_{Ny}$, the Rayleigh resolution $1/T$, the number of frequencies found via prewhitening, and the number of unique frequencies in Table \ref{tab:prewhitening_summary}. The unique frequencies, amplitudes, and phases (as well as corresponding formal errors) found for each star are listed in \S\ref{app:A}. 

We note that, while this prewhitening procedure is capable of accurately describing the TESS light curves, significant frequency-dependent astrophysical and instrumental noise exists in these data, and produces spurious coherent structures that may be mistaken for periodicity. Determining the significance of a detected periodic signal under the null hypothesis of such noise is a non-trivial task. While analytic methods exist to estimate significance in the case of white or power-law noise \citep[e.g.,][]{baluev08,vaughan05}, significance tests in the case of a more complex noise model are inconsistent in various subfields in the literature. No suitable physical model for the astrophysical noise, represented in Equation \eqref{eq:rednoise}, has been proposed; therefore, any estimate of the significance of the frequencies found here would be model dependent. However, we do test to determine if the value of the periodogram peaks closest to the recovered frequencies exceed a power corresponding to a false-alarm level of 1\% under the null hypothesis of white noise, as described by \citet{baluev08}, and note the cases where frequencies discussed do not, both in text and in \S\ref{app:A}. 

In the case of the red noise discussed below, \citet{blomme11} present a method of evaluating the significance of frequencies under noise of the form discussed here. After fitting the periodogram with Equation \eqref{eq:rednoise}, we rescale the noise function so that the integral over the frequency range considered is equal to the variance of the flux times $f_{Ny}$. For each frequency, we evaluate:
\begin{equation}
    P(z>Z) = 1-e^{-e^{-0.93Z+\ln{(0.8N)}}}
\end{equation}
which gives the probability that a stochastic noise process would result in an amplitude $z$ that exceeds a threshold
\begin{equation}
    Z(f) = \frac{A_j^2N}{4\sigma_{red}^2}
\end{equation}
where $A_j$ is the amplitude found by prewhitening, $N$ is the number of points in the lightcurve, and $\sigma_{red}$ is the value of the rescaled noise function at the frequency under consideration. A frequency is considered significant under red noise if $P<0.01$. Suspiciously, all frequencies found are significant under this criterion. As we mention above, this result is likely model dependent; it is possible that many of these frequencies (if not all in some cases) are entirely attributable to the noise process.Future TESS sectors will provide insight into the low frequency regime, and allow us to conduct, for example, time-frequency analyses over a much longer baseline to determine the persistence of these frequencies. Finally, we estimate the signal-to-noise ratio ($SNR$) for each frequency, calculated as the ratio of the amplitude to the standard deviation of the lightcurve after prewhitening. While more straightforward, this estimate of the significance of each frequency is also flawed, as no frequency has $SNR > 2$, despite all frequencies included quantitatively improving the overall fit to the lightcurve according to the adopted $BIC$ criterion (Equation \eqref{eq:bic}). Indeed many of these frequencies being visible by eye in the lightcurves in Figure \ref{fig:lcs}. Nevertheless, we include the measured $SNR$ in the tables in \S\ref{app:A}.

\begin{deluxetable*}{lccccc}
\tabletypesize{\footnotesize}
\tablecaption{Summary of light curve characteristics. \label{tab:prewhitening_summary}.}
\tablehead{\colhead{Common Name} & \colhead{$N$} & \colhead{$f_{Ny}$} & \colhead{$1/T$} & \colhead{Number of frequencies} & \colhead{Unique frequencies} \\
\colhead{} & \colhead{} & \colhead{[day$^{-1}$]} & \colhead{[day$^{-1}$]} & \colhead{} &  \colhead{}}
\startdata
S Dor     & 36296 & 322.8060468379108 & 0.0177879072510214 & 71 & 41 \\
HD 269953 & 36418 & 323.89171072083246 & 0.01778794028727421 & 14 & 14 \\
HD 269582 & 18226 & 332.52345927965945 & 0.036490914598590884 & 33 & 25 \\
HD 270046 & 36291 & 322.76203971547 & 0.017787932748165886 & 14 & 10 \\
HD 270111 & 18101 & 324.59880106819276 & 0.03586727083626439 & 7 & 6 \\
HD 269331 &  18279 & 333.49073111685124 & 0.03649094333262405 & 10 & 6 \\
HD 269110 & 36403 & 323.7577400015701 & 0.017787909455610686 & 11 & 7 \\
HD 268687 & 36412 & 323.83746567083887 & 0.017787891882718898 & 64 & 33 \\
\enddata
\end{deluxetable*}

\section{Results}\label{sec:results}

\subsection{Yellow Supergiants}

A 25 $M_\odot$ solar-metallicity will begin its life as an O star; in the Geneva evolutionary tracks \citep{ekstrom12} this star is an O6 dwarf on the zero-age main sequence. After 7 Myr, it has evolved into a B0 supergiant, at which point it crosses the HR diagram in under a Myr to become a RSG. Approximately 500 kyr later, it has evolved bluewards once more to become a Wolf-Rayet star \citep{massey17}. During both rightward and leftward crossings of the HR diagram, the star undergoes an incredibly brief yellow supergiant (YSG) phase. Thus, while the luminosities and effective temperatures of two given YSGs may be identical, their initial masses, ages, and interior structures may be radically different. Signatures of these differences may be imprinted in the TESS lightcurves. While it is likely that a cool YSG that has previously undergone a RSG phase will be accompanied by a dusty envelope, as the star's effective temperature increases, the dust may be photodissociated, which is consistent with the decreasing abundance of circumstellar dust species around increasingly hot evolved massive stars in Table 1 of \citet{waters10}. Therefore, it is possible that variability may be the best or most unambiguous means of distinguishing between rightward- and leftward-moving YSGs Finding leftward-moving YSGs places a valuable upper limit on the initial masses of stars that explode as RSGs before they can become YSGs (a.k.a., ``the red supergiant problem'', e.g., \citealt{smartt09}).

\begin{figure*}[ht!]
\plotone{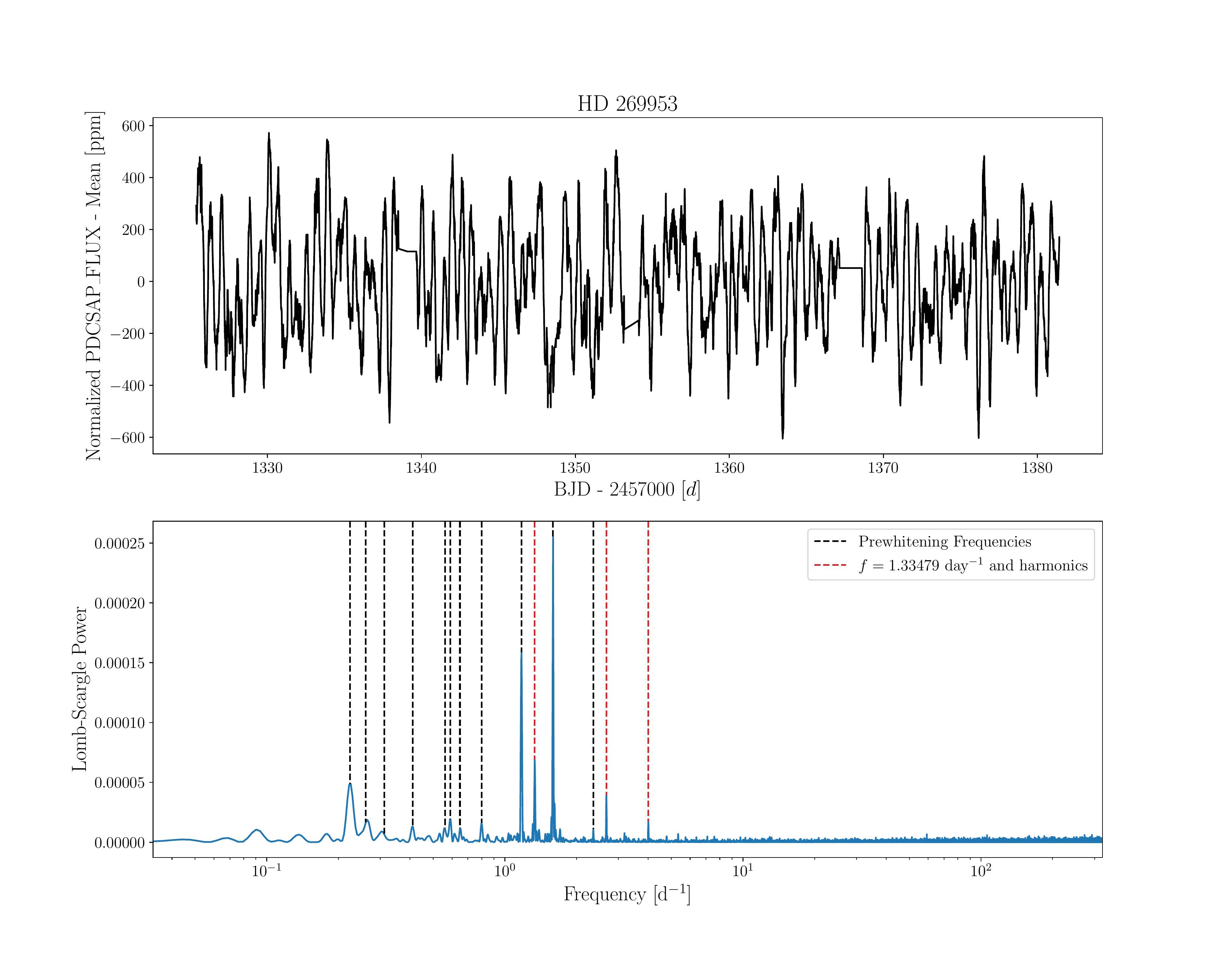}
\caption{{\it Top}: Light curve for HD 269953, after smoothing with a 128-point rolling median, showing coherent variability at approximately 600 ppm. {\it Bottom}: Lomb-Scargle Periodogram. The black lines indicate frequencies found via prewhitening. The $f=1.33479$ day$^{-1}$ peak and it's two detected harmonics are indicated with red vertical lines.}\label{fig:hd269953_lcperiod}
\end{figure*}

\begin{figure*}[ht!]
\plotone{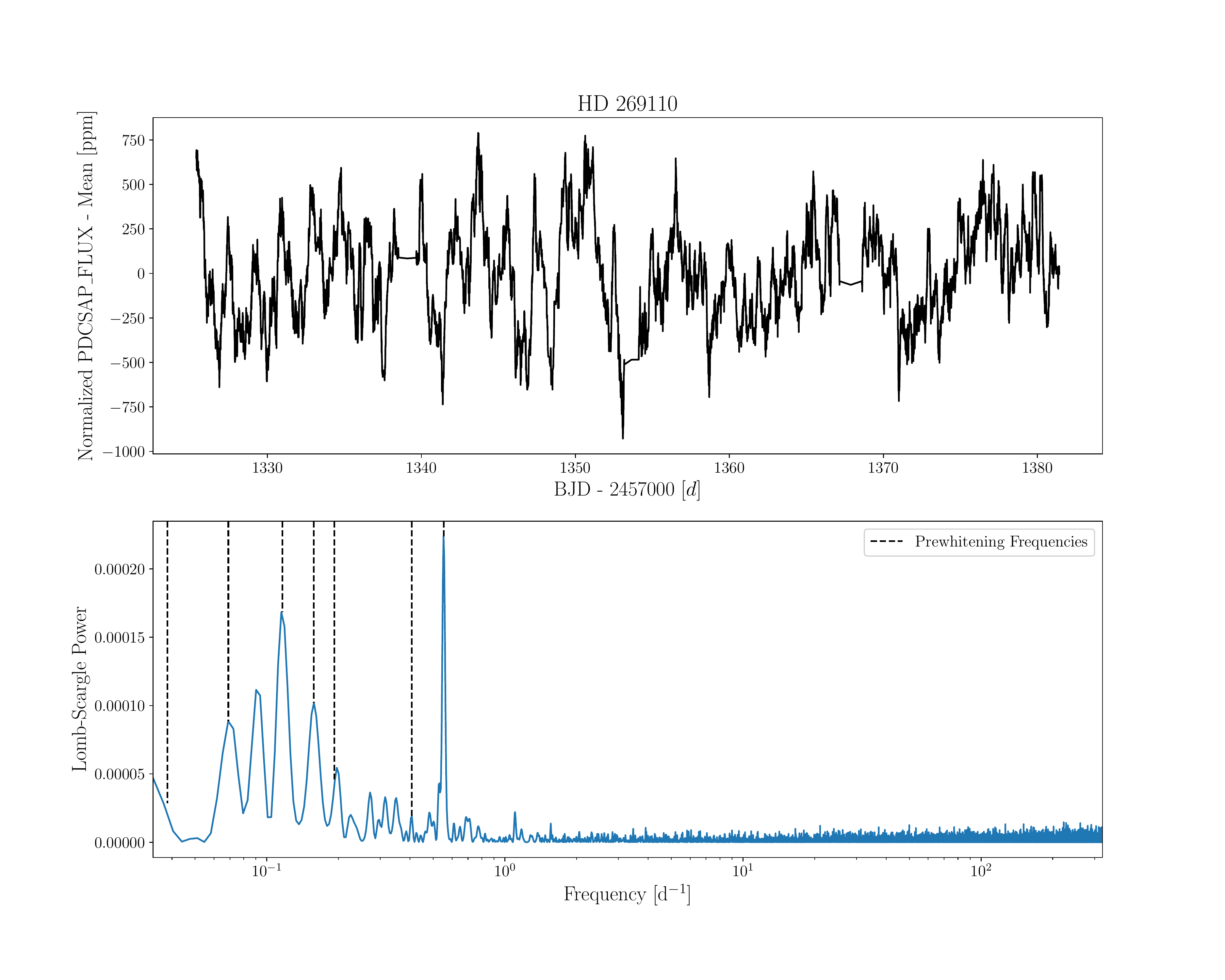}
\caption{Similar to Figure \ref{fig:hd269953_lcperiod}, for HD 269110.}\label{fig:hd269110_lcperiod}
\end{figure*}

\begin{figure*}[ht!]
\plotone{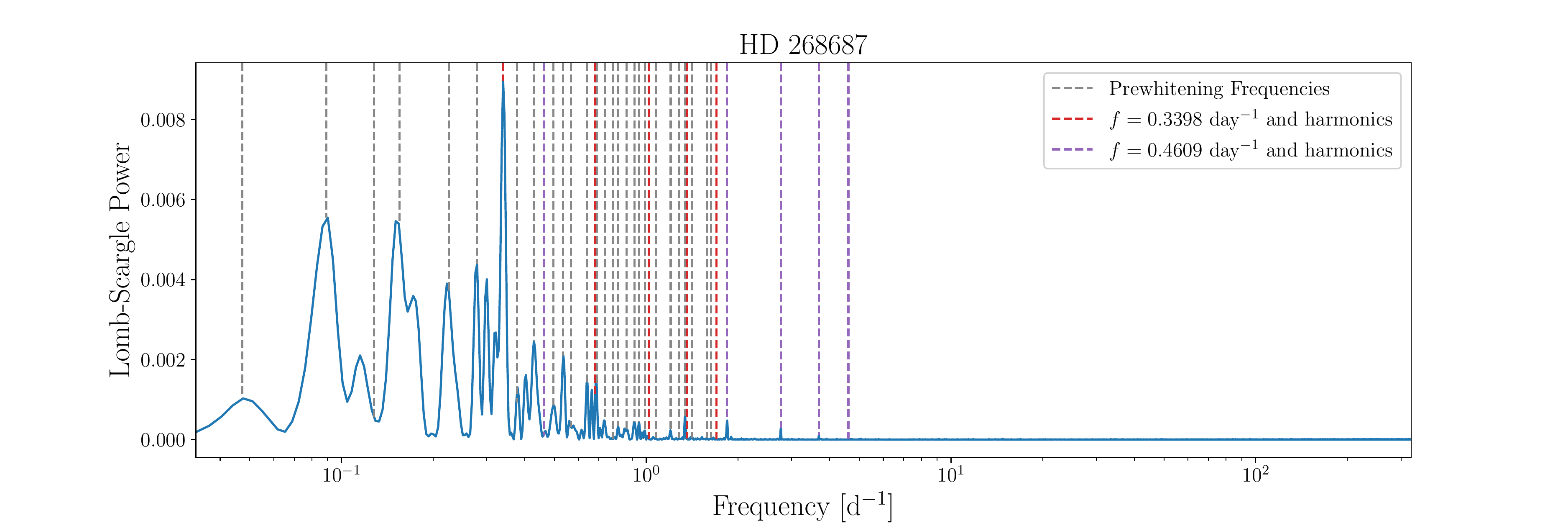}
\caption{Similar to the bottom panel of Figure \ref{fig:hd269953_lcperiod} for HD 268687. We indicate the strongest peak and its first four harmonics with vertical red lines; these harmonics were not detected via the procedure described in \S\ref{subsec:prewhitening}. We also find harmonics of a $\sim0.4609$ day$^{-1}$ signal, indicated in purple.}\label{fig:hd268687_period}
\end{figure*}

\subsubsection{HD 269953}

HD 269953 is a G0 YSG, assigned a luminosity class of 0 by \citet{keenan89}, which is in agreement with its high luminosity $\log(L/L_\odot) = 5.437$ from \citet{neugent12_ysg}. Coupled with its temperature $T_{eff} = 4920$ K, this implies a radius of 566 R$_\odot$ from the Stephan-Boltzmann law. While the light curve presented in Figure \ref{fig:lcs} appears to be dominated by noise, the light curve smoothed by a 128-point rolling median appears to show coherent oscillations. We re-plot the smoothed, mean-subtracted light curve in the top panel of Figure \ref{fig:hd269953_lcperiod}. The periodogram, plotted in the lower panel of Figure \ref{fig:hd269953_lcperiod} shows multiple strong peaks at frequencies above 1 day$^{-1}$. The prewhitening procedure described in \S\ref{subsec:prewhitening} reveals the presence of 14 unique frequencies, indicated by the vertical black lines. We search for harmonics of the form $f_1/f_0$ that satisfy
\begin{equation}
        nf_0 - f_1 \leq \sqrt{\big(n\epsilon(f_0)\big)^2 + \big(\epsilon(f_1)\big)^22}
\end{equation}
where $n$ is an integer greater than 1. In the light curve of HD 269593, we detect the first ($n=2$) and second ($n=3$) harmonics of $f=1.3347924$ day$^{-1}$, which we indicate with the vertical red lines. While we search for harmonics up to $n=10$ for all sources, the low amplitudes and frequency resolution make it unlikely that any high harmonics found are real, and we caution against over-interpreting high harmonics that may not exist. For example, this search also revealed the ninth harmonic of $f=0.26019548$ day$^{-1}$. The corresponding peak in the periodogram does not surpass the 1\% false alarm level; because of this fact, along with the null detection of any of the preceding harmonics, we deem this to be a chance coincidence. 

We also searched for frequency combinations in the form $f_0+f_1=f_2$, such that 
\begin{equation}
        f_0 + f_1 - f_2 \leq \sqrt{\big(\epsilon(f_0)\big)^2+\big(\epsilon(f_1)\big)^2+\big(\epsilon(f_2)\big)^2}
\end{equation}
We find that the strongest peak in the periodogram at $1.59360677$ day$^{-1}$ can be represented as the sum of two peaks at $0.26019548$ and $1.3347924$ day$^{-1}$. If variability at these frequencies is driven by pulsations, this may indicate that some of the modes are interacting with each other.

With high amounts of IR reddening, HD 269953 is the dustiest YSG in our sample, making it quite likely that it is in a post-RSG phase. Thus, a direct measurement of its mass via asteroseismology, using the multiple sets of frequencies presented here would be an incredibly valuable constraint on stellar evolution. Unfortunately, it is located in the star-forming LMC cluster NGC 2085, and thus is subject to a high degree of crowding in TESS's 21'' pixels. While HD 269953 is the brightest star by far in the field, we reserve further analysis of the light curve until more advanced tools are developed to extract light curves from crowded regions in TESS. Regardless, it is readily apparent that the periodogram of HD 269953 is different from the periodograms of the two other YSGs discussed below. The strongest peaks in its periodogram are located at almost an order of magnitude higher frequency, and it displays very little of the low frequency peaks seen in the periodograms of the other YSGs. Additionally, the slope of the background noise (discussed further below) is much flatter. Coupled with its apparently more-evolved state, this suggests a distinct difference in variability between pre- and post-RSG yellow supergiants.

\subsubsection{HD 269110 \& HD 268687}

HD 269110 is a lower luminosity YSG with $\log(L/L_\odot) = 5.251$, $T_{eff} = 5624$ K \citep{neugent12_ysg}, and thus a radius of 445 R$_\odot$. It has a spectral type of G0I from \citet{ardeberg72}. Similar to HD 269953, the light curve presented in Figure \ref{fig:lcs} appears to be just noise, while the light curve smoothed by a 128-point rolling median shows coherent variability at approximately 750 ppm. Figure \ref{fig:hd269110_lcperiod} shows the smoothed, mean-subtracted light curve in the top panel, and the periodogram in the bottom. Prewhitening reveals a set of 7 unique frequencies in the lightcurve. On average, these frequencies are lower than those found in the other YSGs. With physical properties and photometry consistent with all other YSGs other than HD 269953, it is not clear why might be the case; it is possible that, as the second-faintest star in the sample, we simply don't detect higher frequency signals. The strongest peak at 0.55 day$^{-1}$ (a period of 1.81 days) has its first harmonic visible; however, it was not recovered by our prewhitening procedure. \citet{blomme11} found that the BIC was the most conservative stopping criterion in their prewhitening procedure, resulting in the fewest detected frequencies (which may also explain why we do not recover one of the low frequency peaks seen in the periodogram). Alternatively, this harmonic may not be real. Data from future TESS sectors will allow us to more firmly establish or disprove the existence of this harmonic. A group of peaks centered at $\nu_{max}=0.115$ day$^{-1}$ with an average frequency spacing $\Delta\nu=0.032$ day$^{-1}$ is also visible. Similar structures have been found in the periodograms of many pulsating stars, and asteroseismic models have yielded precise measurements of their masses and deep insight into their core structures. Comparable measurements of evolved massive stars would provide a previously inaccessible constraint on a poorly understood phase of massive stellar evolution. However, suitable models of YSG pulsations at these timescales do not yet exist.

HD 268687 is classified as a F6Ia supergiant by \citet{ardeberg72}, and has a luminosity $\log(L/L_\odot) = 5.169$ and effective temperature $T_{eff} = 6081$ K from \citet{neugent12_ysg}, implying a radius of 346 R$_\odot$. The periodogram, shown in Figure \ref{fig:hd268687_period}, with prewhitening frequencies shown in grey, displays a clear peak at 0.3398 day$^{-1}$, corresponding to a period of 2.95 days, which is readily visible in the light curve. We indicate this frequency and its first four harmonics with vertical red lines; none of these exact harmonics are found by prewhitening, however, many of the frequencies recovered are close to the harmonics (though the latter three do not surpass the 1\% false alarm level). Additionally, the series of peaks at $\sim0.3$ day$^{-1}$ appears to repeat with smaller amplitudes at a spacing of $\sim1.2$ day$^{-1}$ until $\sim0.8$ day$^{-1}$, though the exact shape of the peaks changes. The autocorrelation function calculated from the periodogram has a series of peaks with $\Delta f$ between 0.06 and 0.25 day$^{-1}$. A total of 33 unique frequencies are revealed by prewhitening, making its lightcurve the most complex of the YSGs studied.  We do recover the third ($n=4$), fifth, seventh, and ninth harmonics of $f=0.46094861$ day$^{-1}$ (the last of which does not surpass the 1\% false alarm level), which we indicate with vertical purple lines. While the fundamental corresponds to a minimum in the periodogram, the harmonics correspond to the only peaks above $\sim2$ day$^{-1}$. Our search for combinations of frequencies yields two frequencies (0.86314365 and 1.07507415 day$^{-1}$) that mix with the dominant 0.3398 day$^{-1}$ signal, along with 8 other combinations. Finally, a broad bump of peaks accompany the dominant peak at lower frequencies.

All told, both YSGs are in similar physical sates, and both show clear peaks in their periodograms on timescales of 2-3 days, in addition to structure at higher frequencies, and a series of peaks at lower frequencies. We can rule out some possible sources of the dominant signal in both light curves. If both stars are approximately 25 M$_\odot$ and the variability is due to binary interactions with a companion, the companion would have to be approximately 64,000 M$_\odot$ to be in a 2.95-day Keplerian orbit outside of the stellar surface of HD 268687, and 360,000 M$_\odot$ to be in a 1.81-day Keplerian orbit around HD 269110. We determine both scenarios to be highly implausible.

Perhaps the brightness modulations are instead due to one or more spots on the surface of the stars, causing the apparent luminosity to change as the star rotates? If the typical spot latitude were at the stellar equator, then the star would be rotating at approximately 6,000 km s$^{-1}$ for HD 268687, and 12,000 km s$^{-1}$ for HD 269110, well beyond the critical velocity for both stars. However, we cannot rule out a nearly-polar spot. This option is somewhat attractive given the change in the shape of the variability in HD 268687 with time, but would require invoking severe surface differential rotation, as well as extremely fast spot decay times to explain the change of the variability in HD 269110.

The final possibility, which is also consistent with the change in the shape of the variability, is that we are observing coherent pulsational variability in both stars, in addition to the apparent ``frequency comb'' seen in the periodogram of HD 269110. YSGs have been observed to vary with periods of many tens of days \citep{ferro85} caused by He ionization-driven radial pulsations. Perhaps we are observing a very high-order harmonic of a radial mode. Alternately, oscillations in a non-radial mode may be causing this variability. Because both stars are in the LMC, it, and most of our other targets, are in the TESS Continuous Viewing Zone (CVZ), and will be observed almost continuously for a year. If this variability is caused by p- or g-mode pulsations, some of the peaks in the periodogram may resolve into additional frequency combs characteristic of these pulsations. Another option is that the variability is caused by Rossby waves (or r-mode oscillations, \citealt{papaloizou78}), which appear as ``hump and spike'' shapes in the periodogram \citep{saio18}, which have been observed in main sequence F and G stars. While the fundamental mode is located at a slightly lower frequency than the rotational frequency (and hence we run into the same problems as above), higher-azimuthal-order frequencies can arise. Unfortunately the amplitude of the oscillations declines sharply at the higher orders, implying that the rotation speeds would only be a factor of a few slower, which is still physically implausible.

\begin{figure}[ht!]
\includegraphics[width=0.5\textwidth]{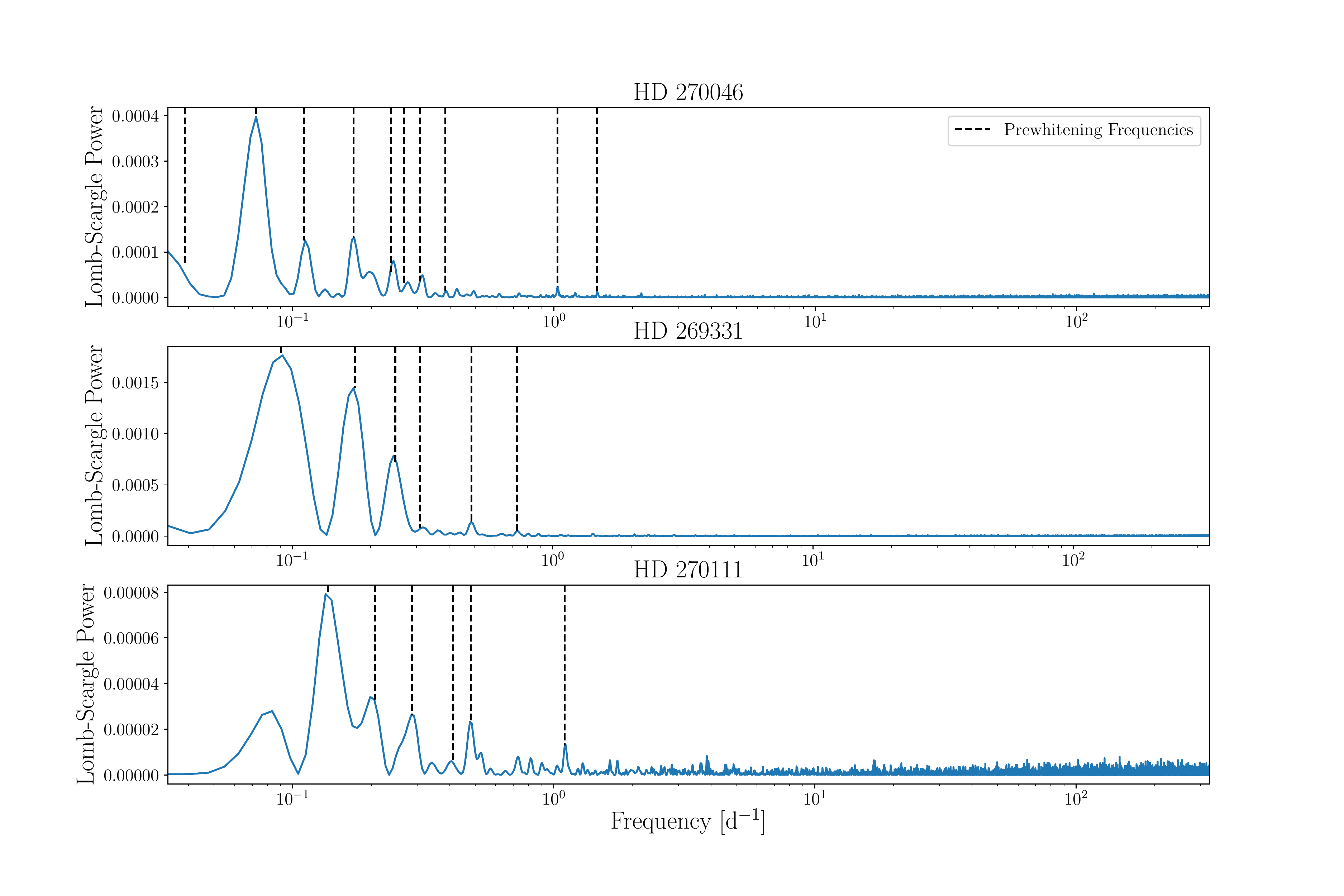}
\caption{Similar to Figure \ref{fig:hd268687_period} for HD 270046, HD 269331, and HD 270111.}
\label{fig:otherysgs_period}
\end{figure}

\begin{figure*}[hp!]
\includegraphics[width=\textwidth]{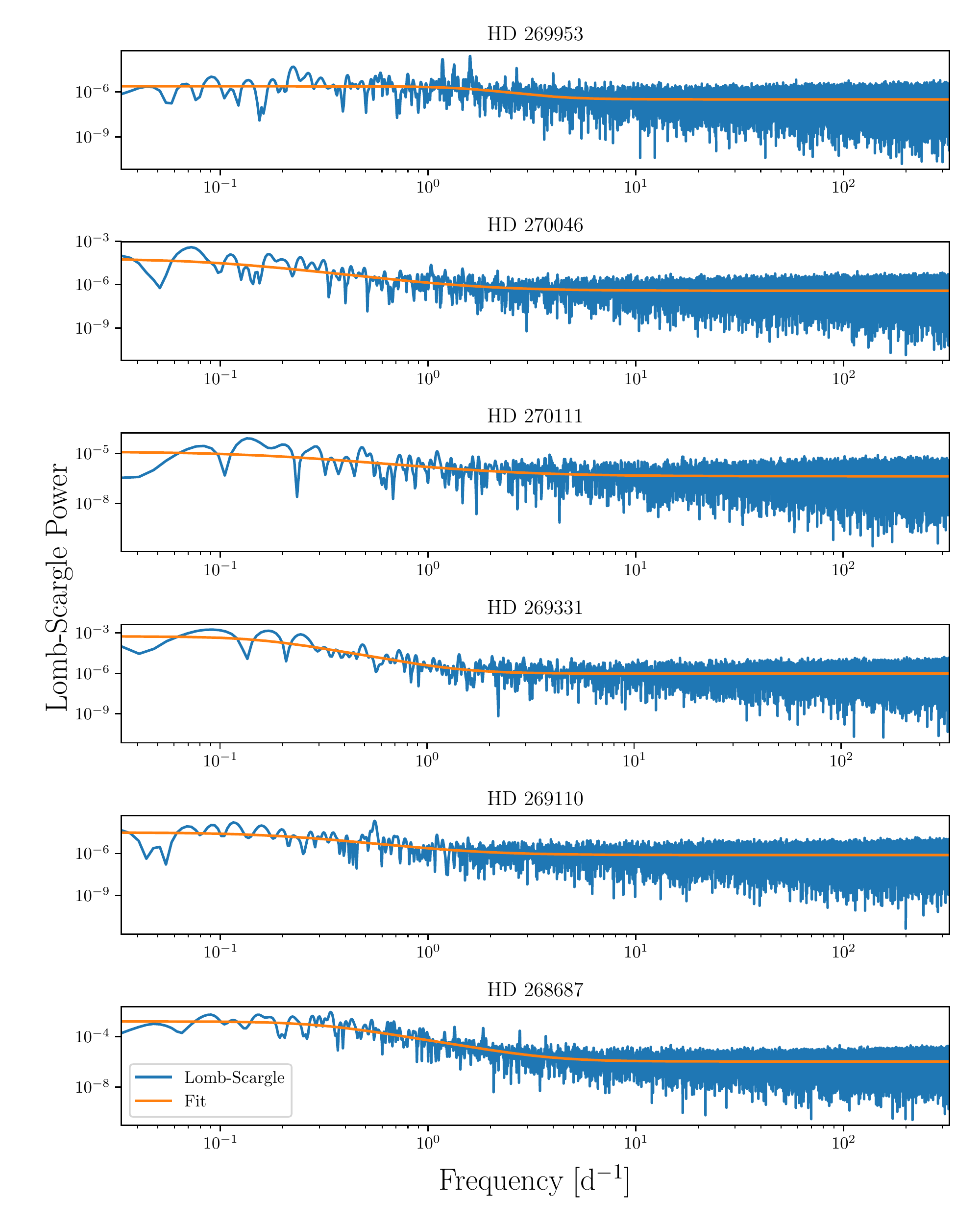}
\caption{Lomb-Scargle periodograms calculated between 1/30 day$^{-1}$ and the pseudo-Nyquist frequency for all six YSGs. Fits using equation \eqref{eq:rednoise} are in orange.}
\label{fig:ysg_noise}
\end{figure*}

\subsubsection{HD 270046, HD 269331, \& HD 270111}

Two of the remaining three YSGs, HD 270046 and HD 270111, have been poorly studied thus far, and display minimal variability in these data. The 5th data release of the RAdial Velocity Experiment (RAVE, \citealt{kunder17}) contains data for HD 270046. While the multiple optical spectra obtained by RAVE yield a wide range of atmospheric parameters (and distances inconsistent with HD 270046's membership in the LMC), measurements of $T_{eff}$ inferred from infrared flux are all between 6200 and 6360 K, with a mean of 6275 K. We were unable to find atmospheric parameters for HD 270111. However, with spectral types of F8Ia and G5I, and TESS magnitudes/infrared colors consistent with HD 269110 and HD 268687, it is fair to assume both stars are fairly typical yellow supergiants. The final YSG, HD 269331, has atmospheric parameters from \citet{neugent12_ysg}. While its temperature is consistent with the other YSGs in their sample, \citet{ardeberg72} assign a spectral type of A5Ia to HD 269331. Its lightcurve (Figure \ref{fig:lcs}) displays two prominent bumps. However, HD 269331 was not observed by TESS in Sector 1, so we are unable to see if these bumps are repeating patterns. We present the periodograms of all three stars in Figure \ref{fig:otherysgs_period}. 

Prewhitening reveals the presence of 10, 6, and 6 unique frequencies in the lightcurves of HD 270046, HD 269331, and HD 270111 respectively, none of which have a normalized semi-amplitude larger than 0.0002, and no peaks surpass the 1\% false alarm level. Some higher order ($n\geq7$) harmonics of various frequencies are found in HD 270046, which we dismiss as coincidental. We also see a combination of the highest peak at $f=0.07243$ day$^{-1}$ and the lowest frequency detected at 0.03865 day$^{-1}$, seen at 0.11055 day$^{-1}$. However, it appears as if the low frequency detected by prewhitening corresponds to a peak in the periodogram that actually lies below our minimum frequency cutoff. In HD 270111, we detect the second and seventh harmonics of the strongest peak at $f=0.136908$ day$^{-1}$; the second harmonic is also the first harmonic of the small peak at 0.207207 day$^{-1}$. With such small amplitudes relative to the noise in the light curve, we refrain from discussing these patterns until higher signal to noise periodograms are obtainable with future TESS sectors.

\subsubsection{Noise Properties of YSG Light Curves}

In addition to the peaks in periodograms of the three YSGs discussed above, background noise exists for all six YSGs. Is this noise instrumental or astrophysical? Astrophysical red noise is seemingly ubiquitous in the light curves of hot massive stars as discussed in \S\ref{sec:intro}, and thus it would be unsurprising for it to manifest in these cooler stars. When plotted in log-scale, some of the periodograms in Figure \ref{fig:ysg_noise} appear to display red noise (especially HD 268687).  To model the background, we follow \citet{blomme11}, and use \texttt{curve\_fit} to fit the Lomb-Scargle periodogram with the function
\begin{equation}\label{eq:rednoise}
    \alpha(f) = \frac{\alpha_0}{1+(2\pi\tau f)^\gamma}+\alpha_w
\end{equation}
from \citet{stanishev02}, where $f$ is the frequency, $\alpha_0$ is the power as $f\rightarrow0$, $\tau$ is a characteristic timescale, and $\alpha_w$ is an additional parameter we add in to model the white noise floor at the highest frequencies (ostensibly equal to the instrumental noise). We perform this fit after calculating the base-10 logarithm of both the Lomb-Scargle power and the fitting function to avoid artificial weighting of real peaks at high frequencies.

The periodograms and fits for all six YSGs are shown in Figure \ref{fig:ysg_noise}. The parameter values and 1$\sigma$ error estimates are compiled in Table \ref{tab:ysg_noise}, and compared to the physical properties of the stars when available in \citet{neugent12_ysg}. It is immediately clear that the noise characteristics of all of the light curves differ, indicating that the source of the noise is likely astrophysical. HD 269953 is quantitatively different from the other YSGs in all parameters but the white noise component of the fit. Notably the red noise power-law component of the fit is only readily apparent over a narrow range of frequencies $\sim2-4$ day$^{-1}$, but the power law slope is approximately twice as steep as all YSGs but HD 268687. Combined with its status as the dustiest YSG in the sample, it is clear that this object warrants further follow-up in the short-timescale regime. Finally, both of the F supergiants have significantly higher power at the lowest frequencies ($\alpha_0$) especially HD 268687. With a larger sample of YSGs, comparisons between physical quantities and noise parameters will help constrain the origin of this noise, which has not been detected until now.

\subsection{Luminous Blue Variables}

\begin{deluxetable*}{lcllllll}
\tabletypesize{\footnotesize}
\tablecaption{Summary of the fit results to the periodograms of the six YSGs in our sample, along with their physical properties when available from \citet{neugent12_ysg}. The $T_{eff}$ for HD 270046 is the mean value inferred from the infrared flux in \citet{kunder17}.
\label{tab:ysg_noise}.}
\tablehead{\colhead{Common Name} & \colhead{Literature Spectral Type} & \colhead{$\log(L/L_\odot)$} & \colhead{$T_{eff}/{\rm K}$} & \colhead{$\alpha_0/10^{-4}$} &  \colhead{$\tau/10^{-2} {\rm d}$}& \colhead{$\gamma$} & \colhead{$\alpha_w/10^{-5}$}}
\startdata
HD 269953 & G0 0 \citep{keenan89} & 5.437 & 4920 & $0.02\pm0.002$ & $8.79\pm0.75$ & $3.06\pm0.28$ & $0.03\pm0.0001$ \\
HD 270046 & F8Ia \citep{ardeberg72} & --- & 6275 &
$0.66\pm0.306$ & $172.14\pm63.31$ & $1.76\pm0.09$ & $0.04\pm0.0002$ \\
HD 270111 & G5I \citep{sanduleak70} & --- & --- & 
$0.13\pm0.056$ & $84.90\pm42.58$ & $1.40\pm0.13$ & $0.04\pm0.0003$ \\
HD 269331 & A5Ia \citep{ardeberg72} & 5.307 & 6457 &
$5.56\pm2.208$ & $100.80\pm22.10$ & $2.83\pm0.19$ & $0.10\pm0.0006$ \\
HD 269110 & G0I \citep{ardeberg72} & 5.251 & 5624 & $0.32\pm0.093$ & $77.23\pm20.29$ & $1.87\pm0.13$ & $0.08\pm0.0003$ \\
HD 268687 & F6Ia \citep{ardeberg72} & 5.169 & 6081 & $16.03\pm2.940$ & $54.06\pm5.25$ & $2.79\pm0.07$ & $0.11\pm0.0005$ \\
\enddata
\end{deluxetable*}

Arguably one of the least understood stellar evolutionary phases, luminous blue variables (LBVs) are a phenomenological class consisting of extremely luminous stars that show signs of dramatic variability. LBVs are perhaps best characterized by their giant eruptions (such as those famously associated with $\eta$ Carina and P Cygni), bright enough to be mistaken as supernovae. In some cases these "impostor" events are followed by true supernovae on timescales of a few years, as in the case of SN 2009ip, which underwent two outbursts in 2009 and 2010 before potentially undergoing a terminal explosion in 2012, e.g. \citep{mauerhan13,fraser15}. However, LBVs also experience large episodic variations in their effective temperatures on timescales of months to years, known as "S Dor variations". With their bolometric luminosities remaining almost constant, these S Dor variations manifest as horizontal evolution on the Hertzsprung-Russell diagram between their hot and cool states. LBVs also exhibit $\sim$0.1 mag irregular microvariability on timescales of weeks to months \citep{abolmasov11}.

The evolutionary state of LBVs, their status as single or binary stars, and the physical mechanisms driving the S Dor variations are all topics of current debate (see \citealt{smith15,humphreys16,aadland18}, Levesque \& Lamers 2019). One possibility is that pulsations may be important for driving mass loss for S Doradus variability \citep{lovekin14}, and may therefore be observable. Indeed, a simple estimate of the dynamical/free-fall timescale for a typical LBV from \citet{abolmasov11} yields
\begin{equation}
    t_{dyn} \approx 0.6 \bigg(\frac{R_*}{10^{12} {\rm cm}}\bigg)^{3/2} \bigg(\frac{M_*}{100 {\rm M}_\odot}\bigg)^{-1/2} {\rm d}
\end{equation}
and variability on this timescale is easily observable by TESS. However, LBVs tend to be surrounded by a complex and sometimes dusty CSM --- indeed, both LBVs studied here have incredibly red colors in Table \ref{tab:sample} --- so these pulsations may be attenuated and modulated by this intervening material. All told, understanding the short timescale variability of LBVs can offer incredibly valuable insight into the physical state of LBVs and their immediate environments.

\subsubsection{HD 269582}

HD (sometimes HDE) 269582 was observed as a H-rich Ofpe/WN9 or WN10h Wolf-Rayet star as recently as the mid 1990s \citep{crowther97}. However, since 2003, it has entered an outbursting LBV state, rapidly brightening in $V$-band as it cooled to a late-B/early-A spectral type, accompanied by drastic changes in various line profiles \citep{walborn17}. Because HD 269582 appears to be newly entering the LBV phase, studying its variability can be quite instructive. Indeed, a link between light curve structure and outbursts has been proposed for Be stars \citep{huat09,kurtz15}; such a link for LBVs may even be testable with an entire year of observations. 

\begin{figure*}[ht!]
\plotone{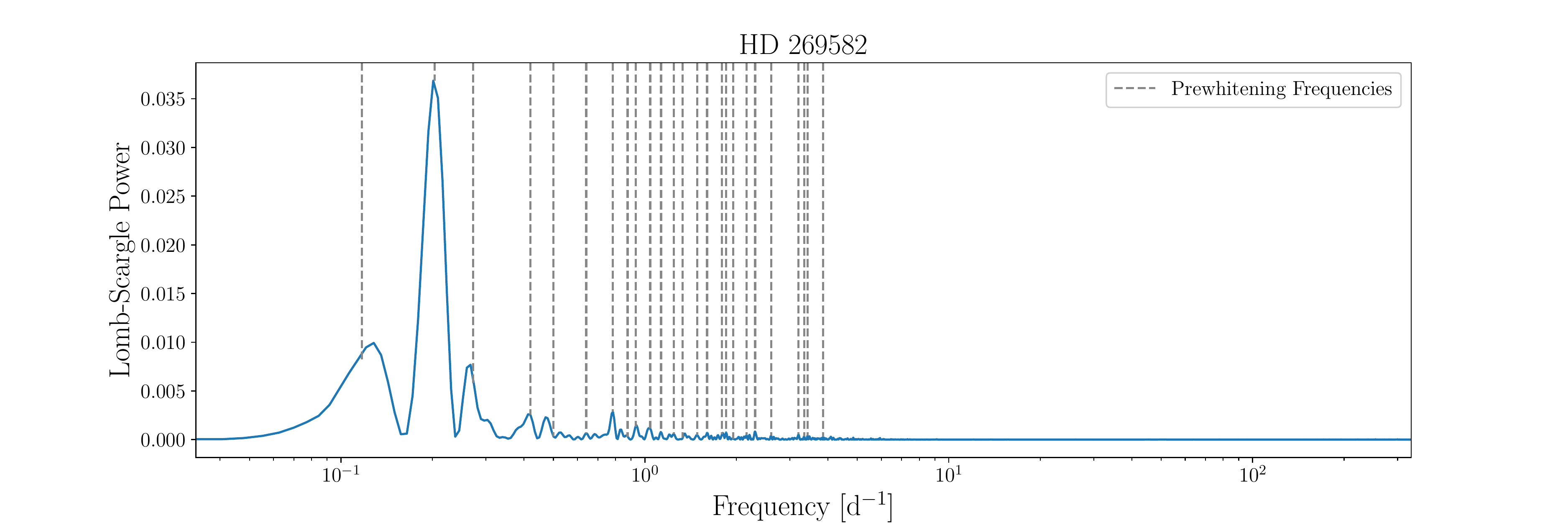}
\caption{Periodogram for HD 269582, showing a clear peak corresponding to a period of 4.919 days. All frequencies found via prewhitening are indicated with grey vertical lines.}\label{fig:hd269582_period}
\end{figure*}

\begin{figure}[ht!]
\includegraphics[width=3.7in]{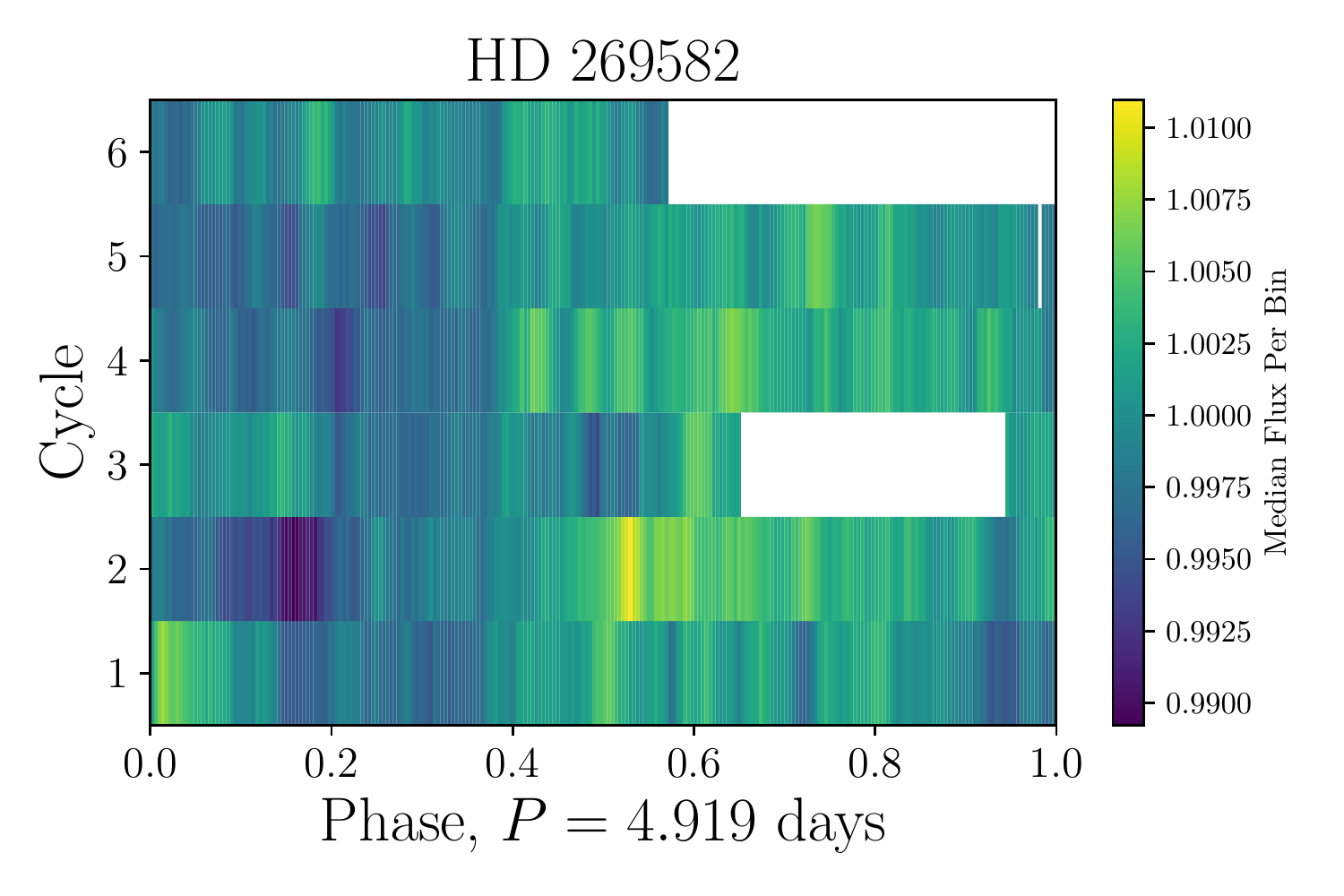}
\caption{Dynamic plot, phased to a 4.919-day period for HD 269582, showing the variability from cycle to cycle.}\label{fig:hd269582_dynamic}
\end{figure}

The light curve for HD 269582 presented in the third panel of Figure \ref{fig:lcs} shows coherent $\sim$1\%-level variability on timescales of a few days. The periodogram shown in Figure \ref{fig:hd269582_period} shows a strong peak, detected with prewhitening at 0.20327588 days$^{-1}$ (corresponding to a 4.919-day period) with small peaks to either side. Though TESS only observed HD 269582 for 5 full cycles of this measured period, the shape of the light curve from cycle to cycle changes noticeably. This can be seen in the dynamic plot in Figure \ref{fig:hd269582_dynamic}, showing the flux as a function of phase from cycle to cycle. The phase of maximum luminosity appears to shift from cycle to cycle, while the amplitude of modulation decreases. Prewhitening reveals the presence of a total of 25 unique frequencies, most of which are small amplitude peaks above $\sim1$ day$^{-1}$. Among those frequencies, we find no convincing harmonics. Interestingly, two frequencies, $f=1.6023258$ and $3.1987555$ day$^{-1}$, are found twice each in our search for sums of frequencies.

Similar dominant periods and changes in the light curve shape were observed in WR 110 by \citet{chene11}. The 30-day light curve presented there appears remarkably similar to the TESS light curve of HD 269582. \citet{chene11} attributed the behavior of WR 110 to a CIR in the wind, implying that we are measuring the rotational frequency. It is also possible that this frequency and the surrounding peaks in the periodogram, or the higher frequencies found by prewhitening are nonradial pulsations. Longer monitoring by TESS will enable us to resolve these peaks further, and build a more physical model with well-sampled parameter distributions, and spectroscopic monitoring would allow us to confirm a CIR scenario.

\subsubsection{S Doradus}

S Doradus is the prototypical S Dor variable, with a long history of photometric and spectroscopic observations. \citet{vangenderen97} detected a $\sim$7 year period in S Dor's light curve, which \citet{abolmasov11} argued is more likely to be a timescale associated with the duration of individual flaring events. 

\begin{figure*}[ht!]
\includegraphics[width=\textwidth]{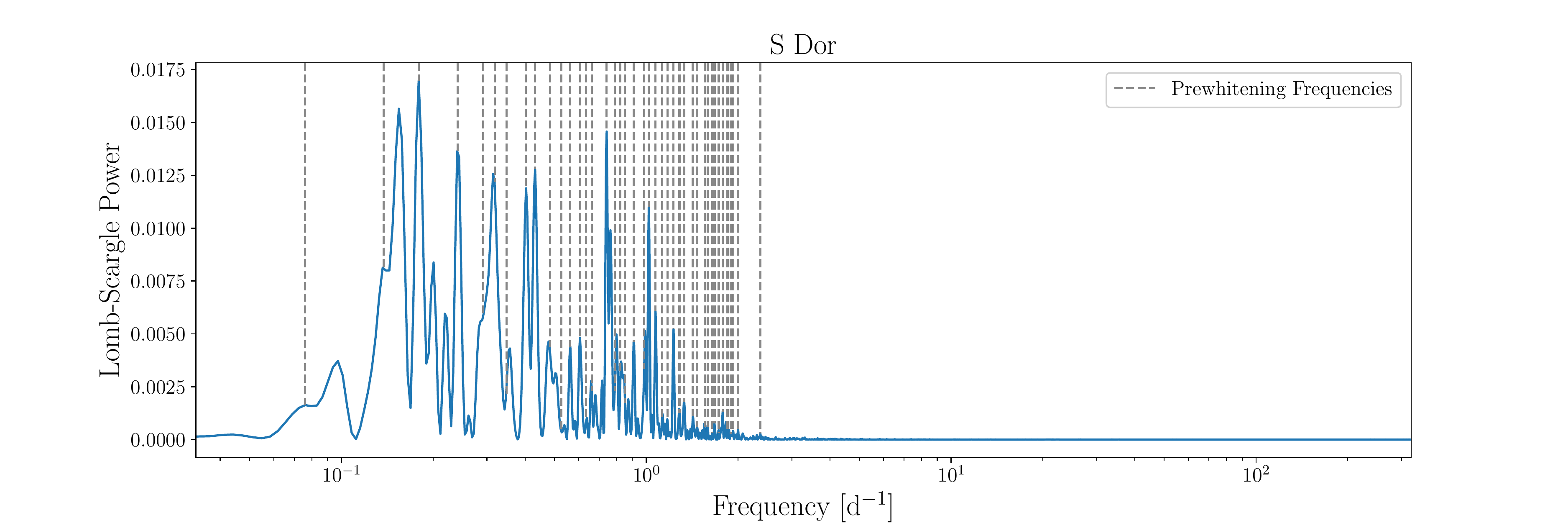}
\caption{Similar to Figure \ref{fig:hd269582_period} for S Dor.}
\label{fig:sdor_period}
\end{figure*}

The light curve presented in Figure \ref{fig:lcs} shows strong $\sim$1\% variations on sub-day timescales. From the periodogram (Figure \ref{fig:sdor_period}), it is clear that the variability displayed by S Dor is quite complicated. Prewhitening reveals a total of 41 unique frequencies, the most of any star in the sample. However, we find no harmonics in this list of frequencies. Similar to HD 269582, we find two frequencies (0.66258409 and 1.6460147 day$^{-1}$) that are each the sum of two different pairs of frequencies. Due to the lack of any single dominant signal, the complexity in the periodogram, and the current theoretical debate on the physical origin of S Dor outbursts, we reserve further modelling until a longer baseline TESS light curve is available, in the hopes of measuring lower frequencies, and resolving the periodogram peaks better.

\begin{figure*}[ht!]
\includegraphics[width=\textwidth]{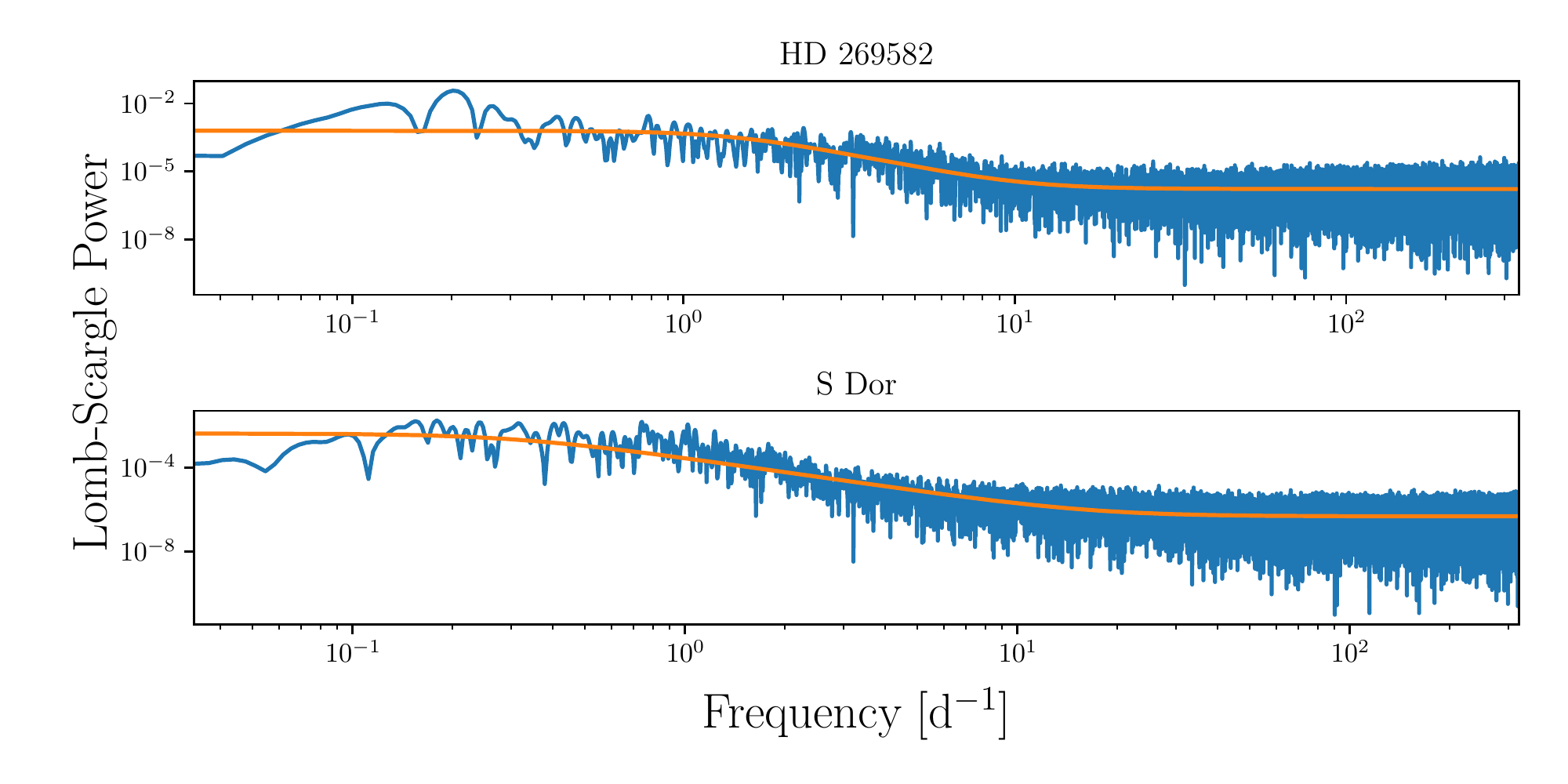}
\caption{Lomb-Scargle periodograms calculated between 1/30 day$^{-1}$ and the pseudo-Nyquist frequency for both LBVs. Fits using equation \eqref{eq:rednoise} are in orange.}
\label{fig:lbv_noise}
\end{figure*}

\subsubsection{LBV Noise Properties}

In addition to our search for coherent variability in the two LBVs, we also analyze the noise properties of their light curves, using Equation \ref{eq:rednoise} to fit the (log of the) Lomb-Scargle periodograms between 1/30 day$^{-1}$ and the pseudo-Nyquist frequency. The resulting fit parameters are presented in Table \ref{tab:lbv_noise}, and the fits themselves are shown in Figure \ref{fig:lbv_noise}. 

The two LBVs have fairly distinct fit properties, as may be expected given their different temperatures, and the recent evolution of HD 269582 into an LBV state. The noise in the S Dor light curve has a higher $\alpha_0$, which may be expected given that the strength of the dominant period in HD 269582. Additionally, we find $\tau\approx0.5$ day and $\gamma=2.29$, while HD 2696582 has $\tau\approx0.1$ and $\gamma=2.96$. In longer-cadence AAVSO data, \citet{abolmasov11} fit the power spectra with a pure power law model, and found slopes closer to 2 for strongly flaring objects, and flatter slopes for LBVs in quiescence. While the TESS data don't probe the low-frequency regime measured by \citet{abolmasov11}, they do indicate that, in HD 268582, the slope of the stochastic noise is steeper than expected, while in S Dor, the region of the power spectrum where the power-law behavior dominates extends over a wide range of frequencies. This suggests that the variability on sub-day timescales in LBVs may be generated by a mixture of physical processes.

\begin{deluxetable}{lllll}
\tabletypesize{\tiny}
\tablecaption{Summary of the fit results to the periodograms of the two LBVs in our sample.\label{tab:lbv_noise}.}
\tablehead{\colhead{Common Name} & \colhead{$\alpha_0/10^{-4}$} &  \colhead{$\tau/10^{-2} {\rm d}$}& \colhead{$\gamma$} & \colhead{$\alpha_w/10^{-5}$}}
\startdata
HD 269582 & $6.33\pm0.74$ & $11.12\pm0.76$ & $2.96\pm0.08$ & $0.17\pm0.0011$ \\
S Dor & $41.23\pm7.10$ & $50.55\pm4.68$ & $2.29\pm0.03$ & $0.05\pm0.000$ \\
\enddata
\end{deluxetable}

\section{Discussion}\label{sec:discuss}

From this small sample of stars it is impossible to make many sweeping inferences. However, the broad range of light curve characteristics, unexpected characteristic time scales, and the structured noise properties displayed by almost every star in this sample make it clear that rare, evolved massive stars are prime candidates for study with TESS and subsequent missions. 

Of the light curves that display clear periodicity, only one (the LBV HD 269582) appears to be on a timescale that could be consistent with a rotational period. Rotation is a deeply important parameter for massive stars which can have drastic effects on their evolution \citep{ekstrom12}. Current samples of measured rotation periods in massive stars are insufficient to statistically measure the distribution of rotation rates, leaving us with spectroscopic measurements \citep[e.g., ][]{huang10} which are hindered by the unknown inclination of the star relative to the line of sight. {\it Kepler} has revolutionized the study of stellar rotation for low-mass (FGKM) stars, increasing the known sample from $\sim10^3$ to over 30,000.
Comparison between physical properties and observed rotation periods for low-mass stars from {\it Kepler} has yielded new insights about magnetic braking evolution and potentially the age distribution of nearby stars in our Galaxy \citep[e.g.][]{van-saders2016,davenport2018}. 

None of the light curves we study display a clear signature of binary interactions. Binary interactions are critical in determining the evolution of many (if not most) massive stars. Galactic O stars have an intrinsic binary fraction of at least $\sim70$\% \citep{sana12}, while the binary fraction in the lower-metallicity LMC appears to be lower \citep{sana13,dornwallenstein18}. Many of the physics governing these interactions can be constrained by observing post-main-sequence massive stars in binary systems. Unfortunately, very few such systems are known: the observed Wolf-Rayet binary fraction is $\sim$30\% \citep{neugent14}, while the binary fraction of yellow and red supergiants is still unknown \citep{levesque17}. It is difficult to reconcile these low numbers with the high binary fraction of main sequence stars. Between the complex circumstellar geometry, and the already-complicated spectra of evolved massive stars, the detection of binary systems via radial velocity measurements is arduous. Photometric diagnostics can be used to find candidate RSG+B systems \citep{neugent18b}, but for many other configurations, photometric variability may be one of the few detectable signatures of binary effects. These variations may manifest themselves as eclipses in well-aligned systems, ellipsoidal variations in short-period systems, or periodic outbursts in extremely eccentric systems as the system approaches periastron. Detecting these effects with TESS and characterizing binary systems with follow-up observations is a critical first step in understanding late-stage massive binary evolution, and resolving the discrepancy between the statistics of main sequence and evolved massive binaries. 

Stars with periodicities inconsistent with rotation or binary interactions possess variability on timescales consistent with pulsations. Pulsational modes can give us deep insight into fundamental stellar properties like mass and radius. However, models of both radial and nonradial pulsations in evolved massive stars have only recently been made available \citep[e.g.,][]{jiang18}. Developing suitable models will allow us to constrain the interior structures of massive stars, and understand energy transport at an unprecedented level. The impact of wave-energy deposition in the last century of a massive star's life can have important impacts on its pre-supernova evolution \citep{fuller17}, and measuring the pulsational properties of the most massive stars will give us valuable constraints on the masses of supernova progenitors.

Finally, red noise is a ubiquitous property in all of the light curves. Whether this noise arises from decoherent pulsations, surface granulations in the cooler stars, wind instabilities in the hot stars, or some other process entirely, measuring the noise characteristics of a large sample of massive stars will allow us to search for trends as a function of evolutionary stage, which can give us some insight into the physical processes involved. All told, studying evolved massive stars at short timescales can help us answer many unsolved problems in massive star evolution. 

\section{Summary \& Conclusion}\label{sec:conclusion}

Our main results are summarized as follows:

\begin{itemize}
  \item We study eight evolved massive stars. We find distinct periodicity in five stars, including two luminous blue variables, and three yellow supergiants. We are unable to constrain the source of the variability in all cases.
  \item The light curve of one YSG, HD 269953, displays unique properties not shared by its fellow YSGs. We suggest that it is in a post-RSG evolutionary phase.
  \item All of the YSGs display red noise in their light curves, that is likely astrophysical in origin.
  \item The LBV HD 269582 displays 1\% variability at $\sim$5 day timescales. While the shape of the variability changes, it is possibly due to a rotation period that is imprinting itself into the wind of HD 269582 via a co-rotating interaction region.
  \item S Doradus exhibits incredible complexity in its peridogram at frequencies below $\sim1.5$ day$^{-1}$, with a total of 41 unique frequencies found via prewhitening.
  \item Both LBVs display red noise. The noise in S Dor is stronger (as parametrized by $\alpha_0$), less steep, and has a longer characteristic timescale $\tau$ compared to HD 269582.
\end{itemize}

We wish to emphasize that evolved massive stars have never been studied before with high cadence space based photometry. As the observed baseline increases for stars in the TESS sourthern CVZ, the periodogram peaks will grow sharper, and allow us to probe lower frequencies for comparison with previous studies. However, our tentative results presented here highlight the importance of studying massive stars in this domain. It is clear that new models are required to explain the observed variability, which will allow these data to give us an incredibly deep insight into the physics of evolved massive stars.

\acknowledgments

This work was supported by NSF grant AST 1714285 awarded to E.M.L.

JRAD acknowledges support from the DIRAC Institute in the Department of Astronomy at the University of Washington. The DIRAC Institute is supported through generous gifts from the Charles and Lisa Simonyi Fund for Arts and Sciences, and the Washington Research Foundation

This work made use of the following software:

\vspace{5mm}

\software{Astropy v3.0.4 \citep{astropy13,astropy18}, Matplotlib v2.2.2 \citep{Hunter:2007}, makecite v0.4 \citep{makecite18}, NumPy v1.15.2 \citep{numpy:2011}, Pandas v0.23.4 \citep{pandas:2010}, Python 3.5.1, SciPy v1.1.0 \citep{scipy:2001}}

\appendix

\section{Frequencies found via prewhitening}\label{app:A}

The following tables contain the list of unique frequencies (separated by $1.5/T$) found by the prewhitening procedure described in \S\ref{subsec:prewhitening}.

\begin{deluxetable*}{cccccccc}
\tabletypesize{\footnotesize}
\tablecaption{Unique frequencies, amplitudes, phases, and formal errors for  S Dor found via prewhitening. For each frequency, we specify the signal to noise as defined in text, and whether the corresponding signal is significant under the assumption of white noise.\label{tab: S Dor_freqs}}
\tablehead{\colhead{$f_j$} & \colhead{$\epsilon(f_j)$} & \colhead{$A_j$} & \colhead{$\epsilon(A_j)$} & \colhead{$\phi_j$} &  \colhead{$\epsilon(\phi_j)$} & \colhead{SNR} & \colhead{White Noise Significant?}\\
\colhead{[day$^{-1}$]} & \colhead{[day$^{-1}$]} & \colhead{[ppt]} & \colhead{[ppt]} & \colhead{[radians]} & \colhead{[radians]} & \colhead{} & \colhead{} } 
\startdata
$0.17914669$ & $0.00018699$ & $1.37355329$ & $0.02618908$ & $2.1115$ & $0.0191$ & 1.3475 & Y \\ 
$0.74026962$ & $0.00019105$ & $1.29287459$ & $0.02518611$ & $-2.1163$ & $0.0195$ & 1.2684 & Y \\ 
$0.40287251$ & $0.00019943$ & $1.14940943$ & $0.02337377$ & $3.0257$ & $0.0203$ & 1.1276 & Y \\ 
$0.43090953$ & $0.00019300$ & $1.14783830$ & $0.02258911$ & $2.2956$ & $0.0197$ & 1.1261 & Y \\ 
$1.01866464$ & $0.00020589$ & $1.03719972$ & $0.02177483$ & $-2.7873$ & $0.0210$ & 1.0176 & Y \\ 
$0.31848154$ & $0.00019611$ & $1.05415657$ & $0.02108043$ & $1.5970$ & $0.0200$ & 1.0342 & Y \\ 
$0.29162660$ & $0.00017781$ & $1.12253453$ & $0.02035210$ & $-2.9533$ & $0.0181$ & 1.1013 & Y \\ 
$0.24027468$ & $0.00021217$ & $0.90005164$ & $0.01947179$ & $0.9284$ & $0.0216$ & 0.8830 & Y \\ 
$0.13744498$ & $0.00021613$ & $0.83184153$ & $0.01833204$ & $-1.7119$ & $0.0220$ & 0.8161 & Y \\ 
$1.07115176$ & $0.00022219$ & $0.78722343$ & $0.01783562$ & $1.8799$ & $0.0227$ & 0.7723 & Y \\ 
$0.78858490$ & $0.00022709$ & $0.74934953$ & $0.01735188$ & $-1.3781$ & $0.0232$ & 0.7352 & Y \\ 
$0.90828327$ & $0.00021351$ & $0.77605388$ & $0.01689571$ & $1.7982$ & $0.0218$ & 0.7614 & Y \\ 
$0.48307755$ & $0.00020698$ & $0.77700877$ & $0.01639875$ & $-2.3994$ & $0.0211$ & 0.7623 & Y \\ 
$0.34807760$ & $0.00022384$ & $0.69613397$ & $0.01588865$ & $2.3071$ & $0.0228$ & 0.6830 & Y \\ 
$0.85068961$ & $0.00022519$ & $0.67309428$ & $0.01545560$ & $-1.3659$ & $0.0230$ & 0.6603 & Y \\ 
$0.56272556$ & $0.00022740$ & $0.64883341$ & $0.01504492$ & $-2.4447$ & $0.0232$ & 0.6365 & Y \\ 
$1.22788836$ & $0.00022903$ & $0.62763187$ & $0.01465752$ & $-0.5198$ & $0.0234$ & 0.6157 & Y \\ 
$0.98355185$ & $0.00025223$ & $0.54339970$ & $0.01397567$ & $-2.0356$ & $0.0257$ & 0.5331 & Y \\ 
$0.07590602$ & $0.00025525$ & $0.52566523$ & $0.01368190$ & $-1.4919$ & $0.0260$ & 0.5157 & Y \\ 
$0.60660139$ & $0.00025350$ & $0.51849540$ & $0.01340258$ & $0.4052$ & $0.0258$ & 0.5087 & Y \\ 
$1.33015001$ & $0.00024735$ & $0.47928094$ & $0.01208818$ & $1.8572$ & $0.0252$ & 0.4702 & Y \\ 
$0.52554769$ & $0.00029181$ & $0.39080118$ & $0.01162824$ & $2.6079$ & $0.0298$ & 0.3834 & Y \\ 
$0.63478994$ & $0.00028666$ & $0.38523180$ & $0.01126049$ & $0.3253$ & $0.0292$ & 0.3779 & Y \\ 
$1.12676389$ & $0.00031689$ & $0.33764956$ & $0.01091027$ & $-2.3219$ & $0.0323$ & 0.3313 & Y \\ 
$1.77927887$ & $0.00032095$ & $0.32894596$ & $0.01076545$ & $1.6639$ & $0.0327$ & 0.3227 & Y \\ 
$1.46672799$ & $0.00031496$ & $0.33088598$ & $0.01062659$ & $-2.3816$ & $0.0321$ & 0.3246 & Y \\ 
$1.42147429$ & $0.00035459$ & $0.27557523$ & $0.00996403$ & $-1.6050$ & $0.0362$ & 0.2704 & Y \\ 
$1.28479399$ & $0.00035017$ & $0.27609702$ & $0.00985824$ & $-2.3815$ & $0.0357$ & 0.2709 & Y \\ 
$0.66258409$ & $0.00034800$ & $0.27480157$ & $0.00975139$ & $-0.1970$ & $0.0355$ & 0.2696 & Y \\ 
$1.67496645$ & $0.00035948$ & $0.26019401$ & $0.00953745$ & $2.8436$ & $0.0367$ & 0.2553 & Y \\ 
$1.59025206$ & $0.00034506$ & $0.26826890$ & $0.00943897$ & $-1.8810$ & $0.0352$ & 0.2632 & Y \\ 
$1.72867338$ & $0.00034824$ & $0.26282574$ & $0.00933267$ & $0.6700$ & $0.0355$ & 0.2578 & Y \\ 
$1.89236098$ & $0.00034930$ & $0.25913313$ & $0.00922956$ & $2.8694$ & $0.0356$ & 0.2542 & Y \\ 
$1.17357337$ & $0.00038477$ & $0.22163510$ & $0.00869573$ & $1.7063$ & $0.0392$ & 0.2174 & Y \\ 
$0.82216092$ & $0.00040298$ & $0.20429334$ & $0.00839472$ & $-1.6606$ & $0.0411$ & 0.2004 & Y \\ 
$1.84679016$ & $0.00041217$ & $0.19482554$ & $0.00818812$ & $0.8427$ & $0.0420$ & 0.1911 & Y \\ 
$1.99892588$ & $0.00039723$ & $0.20056820$ & $0.00812403$ & $0.4310$ & $0.0405$ & 0.1968 & Y \\ 
$1.64601470$ & $0.00039939$ & $0.19780379$ & $0.00805566$ & $-0.7468$ & $0.0407$ & 0.1941 & Y \\ 
$1.92956387$ & $0.00041821$ & $0.18733961$ & $0.00798889$ & $-0.3576$ & $0.0426$ & 0.1838 & Y \\ 
$1.55320350$ & $0.00045138$ & $0.16847434$ & $0.00775430$ & $-0.8069$ & $0.0460$ & 0.1653 & Y \\ 
$2.36401361$ & $0.00048031$ & $0.15626593$ & $0.00765330$ & $1.8377$ & $0.0490$ & 0.1533 & Y 
\enddata
\end{deluxetable*}

\begin{deluxetable*}{cccccccc}
\tabletypesize{\footnotesize}
\tablecaption{Unique frequencies, amplitudes, phases, and formal errors for HD 269953 found via prewhitening. For each frequency, we specify the signal to noise as defined in text, and whether the corresponding signal is significant under the assumption of white noise.\label{tab:HD 269953_freqs}}
\tablehead{\colhead{$f_j$} & \colhead{$\epsilon(f_j)$} & \colhead{$A_j$} & \colhead{$\epsilon(A_j)$} & \colhead{$\phi_j$} &  \colhead{$\epsilon(\phi_j)$} & \colhead{SNR} & \colhead{White Noise Significant?}\\
\colhead{[day$^{-1}$]} & \colhead{[day$^{-1}$]} & \colhead{[ppt]} & \colhead{[ppt]} & \colhead{[radians]} & \colhead{[radians]} & \colhead{} & \colhead{} } 
\startdata
$1.59360677$ & $0.00033360$ & $0.16933302$ & $0.00576003$ & $-2.4637$ & $0.0340$ & 0.2248 & Y \\ 
$1.17415786$ & $0.00045342$ & $0.12308820$ & $0.00569092$ & $-0.7787$ & $0.0462$ & 0.1634 & Y \\ 
$1.33479243$ & $0.00067704$ & $0.08190471$ & $0.00565440$ & $-1.2113$ & $0.0690$ & 0.1087 & Y \\ 
$0.22397835$ & $0.00077286$ & $0.07154318$ & $0.00563809$ & $2.3700$ & $0.0788$ & 0.0950 & Y \\ 
$2.67017901$ & $0.00084583$ & $0.06522685$ & $0.00562564$ & $0.3010$ & $0.0862$ & 0.0866 & Y \\ 
$0.58964787$ & $0.00123743$ & $0.04450296$ & $0.00561529$ & $-1.1783$ & $0.1262$ & 0.0591 & Y \\ 
$4.00557547$ & $0.00130137$ & $0.04228006$ & $0.00561047$ & $1.8289$ & $0.1327$ & 0.0561 & Y \\ 
$0.79888166$ & $0.00134452$ & $0.04089120$ & $0.00560608$ & $2.3207$ & $0.1371$ & 0.0543 & Y \\ 
$2.35305773$ & $0.00136719$ & $0.04018379$ & $0.00560199$ & $-2.6328$ & $0.1394$ & 0.0533 & Y \\ 
$0.56055207$ & $0.00141498$ & $0.03879914$ & $0.00559803$ & $0.5170$ & $0.1443$ & 0.0515 & Y \\ 
$0.26019548$ & $0.00145392$ & $0.03773483$ & $0.00559432$ & $1.0507$ & $0.1483$ & 0.0501 & Y \\ 
$0.41029670$ & $0.00166929$ & $0.03284581$ & $0.00559081$ & $2.5195$ & $0.1702$ & 0.0436 & Y \\ 
$0.64779043$ & $0.00171080$ & $0.03203336$ & $0.00558813$ & $-3.1395$ & $0.1744$ & 0.0425 & Y \\ 
$0.31217690$ & $0.00180051$ & $0.03042365$ & $0.00558561$ & $0.1242$ & $0.1836$ & 0.0404 & N 
\enddata
\end{deluxetable*}

\begin{deluxetable*}{cccccccc}
\tabletypesize{\footnotesize}
\tablecaption{Unique frequencies, amplitudes, phases, and formal errors for HD 269582 found via prewhitening. For each frequency, we specify the signal to noise as defined in text, and whether the corresponding signal is significant under the assumption of white noise.\label{tab:HD 269582_freqs}}
\tablehead{\colhead{$f_j$} & \colhead{$\epsilon(f_j)$} & \colhead{$A_j$} & \colhead{$\epsilon(A_j)$} & \colhead{$\phi_j$} &  \colhead{$\epsilon(\phi_j)$} & \colhead{SNR} & \colhead{White Noise Significant?}\\
\colhead{[day$^{-1}$]} & \colhead{[day$^{-1}$]} & \colhead{[ppt]} & \colhead{[ppt]} & \colhead{[radians]} & \colhead{[radians]} & \colhead{} & \colhead{} } 
\startdata
$0.20327588$ & $0.00026040$ & $2.82300142$ & $0.03653911$ & $1.9457$ & $0.0129$ & 1.4285 & Y \\ 
$0.11705849$ & $0.00038216$ & $1.56893390$ & $0.02980298$ & $-1.8281$ & $0.0190$ & 0.7939 & Y \\ 
$0.27226816$ & $0.00053917$ & $1.02731918$ & $0.02753193$ & $-2.3371$ & $0.0268$ & 0.5199 & Y \\ 
$0.78368008$ & $0.00069788$ & $0.76229473$ & $0.02644290$ & $0.0658$ & $0.0347$ & 0.3857 & Y \\ 
$0.42011270$ & $0.00076838$ & $0.67630930$ & $0.02583021$ & $1.8420$ & $0.0382$ & 0.3422 & Y \\ 
$0.93338470$ & $0.00082913$ & $0.59356109$ & $0.02446216$ & $2.5364$ & $0.0412$ & 0.3004 & Y \\ 
$1.04145053$ & $0.00111576$ & $0.42869614$ & $0.02377533$ & $-1.8655$ & $0.0555$ & 0.2169 & Y \\ 
$1.60232584$ & $0.00110823$ & $0.42783042$ & $0.02356714$ & $-2.4842$ & $0.0551$ & 0.2165 & Y \\ 
$2.30579937$ & $0.00112019$ & $0.41940449$ & $0.02335219$ & $1.7965$ & $0.0557$ & 0.2122 & Y \\ 
$1.84981762$ & $0.00115772$ & $0.39499834$ & $0.02273024$ & $1.7261$ & $0.0575$ & 0.1999 & Y \\ 
$3.19875548$ & $0.00124431$ & $0.36446275$ & $0.02254161$ & $-0.4055$ & $0.0618$ & 0.1844 & Y \\ 
$1.24393495$ & $0.00124528$ & $0.36155380$ & $0.02237916$ & $-2.3360$ & $0.0619$ & 0.1830 & Y \\ 
$0.64113112$ & $0.00136209$ & $0.32818766$ & $0.02221935$ & $-0.2110$ & $0.0677$ & 0.1661 & Y \\ 
$2.60327615$ & $0.00142004$ & $0.31293170$ & $0.02208795$ & $-2.7382$ & $0.0706$ & 0.1584 & Y \\ 
$2.15944146$ & $0.00143368$ & $0.30823824$ & $0.02196565$ & $-1.2146$ & $0.0713$ & 0.1560 & Y \\ 
$1.48653861$ & $0.00144310$ & $0.30457385$ & $0.02184704$ & $3.0061$ & $0.0717$ & 0.1541 & Y \\ 
$0.87749148$ & $0.00150520$ & $0.28886982$ & $0.02161222$ & $-2.1677$ & $0.0748$ & 0.1462 & Y \\ 
$3.42963659$ & $0.00152746$ & $0.28324793$ & $0.02150505$ & $-1.3241$ & $0.0759$ & 0.1433 & Y \\ 
$1.79273050$ & $0.00169152$ & $0.25455605$ & $0.02140248$ & $-0.6712$ & $0.0841$ & 0.1288 & Y \\ 
$1.95391049$ & $0.00168754$ & $0.25415661$ & $0.02131865$ & $-1.6620$ & $0.0839$ & 0.1286 & N \\ 
$1.33240989$ & $0.00234358$ & $0.18158923$ & $0.02115312$ & $-0.4892$ & $0.1165$ & 0.0919 & N \\ 
$1.13080040$ & $0.00161657$ & $0.26272929$ & $0.02111096$ & $-2.4513$ & $0.0804$ & 0.1329 & Y \\ 
$0.50022915$ & $0.00173592$ & $0.24363786$ & $0.02102219$ & $-0.9353$ & $0.0863$ & 0.1233 & Y \\ 
$3.86015782$ & $0.00167586$ & $0.25044910$ & $0.02086229$ & $-1.1760$ & $0.0833$ & 0.1267 & Y \\ 
$3.34708610$ & $0.00170259$ & $0.24554829$ & $0.02078023$ & $-2.5729$ & $0.0846$ & 0.1243 & Y 
\enddata
\end{deluxetable*}

\begin{deluxetable*}{cccccccc}
\tabletypesize{\footnotesize}
\tablecaption{Unique frequencies, amplitudes, phases, and formal errors for HD 270046 found via prewhitening. For each frequency, we specify the signal to noise as defined in text, and whether the corresponding signal is significant under the assumption of white noise.\label{tab:HD 270046_freqs}}
\tablehead{\colhead{$f_j$} & \colhead{$\epsilon(f_j)$} & \colhead{$A_j$} & \colhead{$\epsilon(A_j)$} & \colhead{$\phi_j$} &  \colhead{$\epsilon(\phi_j)$} & \colhead{SNR} & \colhead{White Noise Significant?}\\
\colhead{[day$^{-1}$]} & \colhead{[day$^{-1}$]} & \colhead{[ppt]} & \colhead{[ppt]} & \colhead{[radians]} & \colhead{[radians]} & \colhead{} & \colhead{} } 
\startdata
$0.07243476$ & $0.00030333$ & $0.20507787$ & $0.00634302$ & $-0.3777$ & $0.0309$ & 0.2496 & Y \\ 
$0.17088628$ & $0.00053075$ & $0.11542925$ & $0.00624698$ & $1.5664$ & $0.0541$ & 0.1405 & Y \\ 
$0.11054852$ & $0.00062375$ & $0.09772997$ & $0.00621592$ & $2.9262$ & $0.0636$ & 0.1189 & Y \\ 
$0.03864776$ & $0.00059915$ & $0.10138306$ & $0.00619388$ & $-1.7175$ & $0.0611$ & 0.1234 & Y \\ 
$0.23757177$ & $0.00081819$ & $0.07373290$ & $0.00615146$ & $-0.3369$ & $0.0834$ & 0.0897 & Y \\ 
$0.30727002$ & $0.00088671$ & $0.06790027$ & $0.00613926$ & $-2.1785$ & $0.0904$ & 0.0826 & Y \\ 
$0.26646463$ & $0.00108406$ & $0.05544526$ & $0.00612890$ & $-1.8481$ & $0.1105$ & 0.0675 & Y \\ 
$1.03232899$ & $0.00124699$ & $0.04814869$ & $0.00612224$ & $-2.4317$ & $0.1272$ & 0.0586 & Y \\ 
$0.38422638$ & $0.00153431$ & $0.03905225$ & $0.00610975$ & $2.7074$ & $0.1565$ & 0.0475 & Y \\ 
$1.46213112$ & $0.00165260$ & $0.03621682$ & $0.00610298$ & $-1.3304$ & $0.1685$ & 0.0441 & Y 
\enddata
\end{deluxetable*}

\begin{deluxetable*}{cccccccc}
\tabletypesize{\footnotesize}
\tablecaption{Unique frequencies, amplitudes, phases, and formal errors for HD 270111 found via prewhitening. For each frequency, we specify the signal to noise as defined in text, and whether the corresponding signal is significant under the assumption of white noise.\label{tab:HD 270111_freqs}}
\tablehead{\colhead{$f_j$} & \colhead{$\epsilon(f_j)$} & \colhead{$A_j$} & \colhead{$\epsilon(A_j)$} & \colhead{$\phi_j$} &  \colhead{$\epsilon(\phi_j)$} & \colhead{SNR} & \colhead{White Noise Significant?}\\
\colhead{[day$^{-1}$]} & \colhead{[day$^{-1}$]} & \colhead{[ppt]} & \colhead{[ppt]} & \colhead{[radians]} & \colhead{[radians]} & \colhead{} & \colhead{} } 
\startdata
$0.13690821$ & $0.00134854$ & $0.13655040$ & $0.00931212$ & $2.5988$ & $0.0682$ & 0.1563 & Y \\ 
$0.20720738$ & $0.00228811$ & $0.08001971$ & $0.00925903$ & $-2.9088$ & $0.1157$ & 0.0916 & Y \\ 
$0.48123388$ & $0.00267880$ & $0.06819859$ & $0.00923862$ & $-2.4779$ & $0.1355$ & 0.0781 & Y \\ 
$0.28689396$ & $0.00253835$ & $0.07186560$ & $0.00922492$ & $-1.0078$ & $0.1284$ & 0.0823 & Y \\ 
$0.41136823$ & $0.00327074$ & $0.05561540$ & $0.00919881$ & $0.6688$ & $0.1654$ & 0.0637 & N \\ 
$1.10220750$ & $0.00358087$ & $0.05074735$ & $0.00918952$ & $2.2761$ & $0.1811$ & 0.0581 & N 
\enddata
\end{deluxetable*}

\begin{deluxetable*}{cccccccc}
\tabletypesize{\footnotesize}
\tablecaption{Unique frequencies, amplitudes, phases, and formal errors for HD 269331 found via prewhitening. For each frequency, we specify the signal to noise as defined in text, and whether the corresponding signal is significant under the assumption of white noise.\label{tab:HD 269331_freqs}}
\tablehead{\colhead{$f_j$} & \colhead{$\epsilon(f_j)$} & \colhead{$A_j$} & \colhead{$\epsilon(A_j)$} & \colhead{$\phi_j$} &  \colhead{$\epsilon(\phi_j)$} & \colhead{SNR} & \colhead{White Noise Significant?}\\
\colhead{[day$^{-1}$]} & \colhead{[day$^{-1}$]} & \colhead{[ppt]} & \colhead{[ppt]} & \colhead{[radians]} & \colhead{[radians]} & \colhead{} & \colhead{} } 
\startdata
$0.09042104$ & $0.00050776$ & $0.61699402$ & $0.01557183$ & $-0.4965$ & $0.0252$ & 0.4745 & Y \\ 
$0.17425772$ & $0.00049282$ & $0.60733804$ & $0.01487734$ & $-2.6545$ & $0.0245$ & 0.4670 & Y \\ 
$0.24876181$ & $0.00072428$ & $0.39309944$ & $0.01415176$ & $-0.3416$ & $0.0360$ & 0.3023 & Y \\ 
$0.48741510$ & $0.00153999$ & $0.18011694$ & $0.01378726$ & $0.6008$ & $0.0765$ & 0.1385 & Y \\ 
$0.31028888$ & $0.00155887$ & $0.17711953$ & $0.01372399$ & $-2.5111$ & $0.0775$ & 0.1362 & Y \\ 
$0.72997431$ & $0.00315091$ & $0.08710461$ & $0.01364212$ & $-0.4636$ & $0.1566$ & 0.0670 & Y 
\enddata
\end{deluxetable*}

\begin{deluxetable*}{cccccccc}
\tabletypesize{\footnotesize}
\tablecaption{Unique frequencies, amplitudes, phases, and formal errors for HD 269110 found via prewhitening. For each frequency, we specify the signal to noise as defined in text, and whether the corresponding signal is significant under the assumption of white noise.\label{tab:HD 269110_freqs}}
\tablehead{\colhead{$f_j$} & \colhead{$\epsilon(f_j)$} & \colhead{$A_j$} & \colhead{$\epsilon(A_j)$} & \colhead{$\phi_j$} &  \colhead{$\epsilon(\phi_j)$} & \colhead{SNR} & \colhead{White Noise Significant?}\\
\colhead{[day$^{-1}$]} & \colhead{[day$^{-1}$]} & \colhead{[ppt]} & \colhead{[ppt]} & \colhead{[radians]} & \colhead{[radians]} & \colhead{} & \colhead{} } 
\startdata
$0.55347685$ & $0.00054988$ & $0.15765766$ & $0.00883984$ & $-3.0872$ & $0.0561$ & 0.1345 & Y \\ 
$0.11644097$ & $0.00063650$ & $0.13560965$ & $0.00880137$ & $0.0247$ & $0.0649$ & 0.1156 & Y \\ 
$0.15753531$ & $0.00084280$ & $0.10207525$ & $0.00877217$ & $-2.7811$ & $0.0859$ & 0.0871 & Y \\ 
$0.06899622$ & $0.00092764$ & $0.09256332$ & $0.00875550$ & $0.3983$ & $0.0946$ & 0.0789 & Y \\ 
$0.19233848$ & $0.00121647$ & $0.07031901$ & $0.00872248$ & $-2.6869$ & $0.1240$ & 0.0600 & Y \\ 
$0.03827322$ & $0.00123650$ & $0.06911817$ & $0.00871466$ & $-2.0575$ & $0.1261$ & 0.0589 & Y \\ 
$0.40617680$ & $0.00160856$ & $0.05304164$ & $0.00870001$ & $-0.8097$ & $0.1640$ & 0.0452 & N 
\enddata
\end{deluxetable*}

\begin{deluxetable*}{cccccccc}
\tabletypesize{\footnotesize}
\tablecaption{Unique frequencies, amplitudes, phases, and formal errors for HD 268687 found via prewhitening. For each frequency, we specify the signal to noise as defined in text, and whether the corresponding signal is significant under the assumption of white noise.\label{tab:HD 268687_freqs}}
\tablehead{\colhead{$f_j$} & \colhead{$\epsilon(f_j)$} & \colhead{$A_j$} & \colhead{$\epsilon(A_j)$} & \colhead{$\phi_j$} &  \colhead{$\epsilon(\phi_j)$} & \colhead{SNR} & \colhead{White Noise Significant?}\\
\colhead{[day$^{-1}$]} & \colhead{[day$^{-1}$]} & \colhead{[ppt]} & \colhead{[ppt]} & \colhead{[radians]} & \colhead{[radians]} & \colhead{} & \colhead{} } 
\startdata
$0.33981559$ & $0.00016013$ & $0.99396295$ & $0.01623009$ & $2.0925$ & $0.0163$ & 0.7147 & Y \\ 
$0.08920241$ & $0.00018662$ & $0.80772984$ & $0.01537070$ & $1.5642$ & $0.0190$ & 0.5808 & Y \\ 
$0.15510687$ & $0.00020617$ & $0.70324644$ & $0.01478409$ & $-1.6240$ & $0.0210$ & 0.5056 & Y \\ 
$0.27854537$ & $0.00024130$ & $0.58182295$ & $0.01431571$ & $-0.7820$ & $0.0246$ & 0.4183 & Y \\ 
$0.42764474$ & $0.00023359$ & $0.58723924$ & $0.01398757$ & $-0.7239$ & $0.0238$ & 0.4222 & Y \\ 
$0.12802564$ & $0.00026270$ & $0.49788111$ & $0.01333659$ & $1.4333$ & $0.0268$ & 0.3580 & Y \\ 
$0.63934778$ & $0.00026317$ & $0.48760456$ & $0.01308489$ & $-3.0826$ & $0.0268$ & 0.3506 & Y \\ 
$0.04723986$ & $0.00032221$ & $0.37892065$ & $0.01244933$ & $2.3780$ & $0.0329$ & 0.2725 & Y \\ 
$0.22510722$ & $0.00031700$ & $0.38006651$ & $0.01228505$ & $-2.1114$ & $0.0323$ & 0.2733 & Y \\ 
$0.53353809$ & $0.00032444$ & $0.36148395$ & $0.01195896$ & $-2.5173$ & $0.0331$ & 0.2599 & Y \\ 
$0.49649588$ & $0.00039730$ & $0.29148694$ & $0.01180883$ & $-2.3401$ & $0.0405$ & 0.2096 & Y \\ 
$0.68791268$ & $0.00041399$ & $0.27292858$ & $0.01152133$ & $-1.9944$ & $0.0422$ & 0.1962 & Y \\ 
$1.34086866$ & $0.00045333$ & $0.24730899$ & $0.01143192$ & $-2.5120$ & $0.0462$ & 0.1778 & Y \\ 
$0.37663919$ & $0.00048990$ & $0.22604532$ & $0.01129199$ & $-2.2111$ & $0.0500$ & 0.1625 & Y \\ 
$0.91512039$ & $0.00049874$ & $0.21952667$ & $0.01116421$ & $2.4091$ & $0.0509$ & 0.1578 & Y \\ 
$1.84286938$ & $0.00049901$ & $0.21823149$ & $0.01110428$ & $-1.1779$ & $0.0509$ & 0.1569 & Y \\ 
$0.73238778$ & $0.00054246$ & $0.19967911$ & $0.01104494$ & $2.9946$ & $0.0553$ & 0.1436 & Y \\ 
$0.80885741$ & $0.00057409$ & $0.18694716$ & $0.01094361$ & $-1.4790$ & $0.0585$ & 0.1344 & Y \\ 
$0.56637230$ & $0.00058752$ & $0.18116978$ & $0.01085355$ & $-0.1779$ & $0.0599$ & 0.1303 & Y \\ 
$0.94887342$ & $0.00059490$ & $0.17821977$ & $0.01081098$ & $0.5464$ & $0.0607$ & 0.1281 & Y \\ 
$2.76492751$ & $0.00062369$ & $0.16936638$ & $0.01077111$ & $0.0375$ & $0.0636$ & 0.1218 & Y \\ 
$0.86314365$ & $0.00069636$ & $0.15073177$ & $0.01070295$ & $-0.4226$ & $0.0710$ & 0.1084 & Y \\ 
$0.77723411$ & $0.00076226$ & $0.13697958$ & $0.01064685$ & $-1.6051$ & $0.0777$ & 0.0985 & N \\ 
$1.20371478$ & $0.00083481$ & $0.12400270$ & $0.01055564$ & $0.6327$ & $0.0851$ & 0.0892 & Y \\ 
$3.68650176$ & $0.00097373$ & $0.10559870$ & $0.01048488$ & $-0.9334$ & $0.0993$ & 0.0759 & Y \\ 
$0.99151843$ & $0.00101754$ & $0.10078308$ & $0.01045694$ & $2.3637$ & $0.1038$ & 0.0725 & Y \\ 
$0.46094861$ & $0.00124317$ & $0.08212864$ & $0.01041091$ & $-1.0581$ & $0.1268$ & 0.0591 & Y \\ 
$1.28252486$ & $0.00126920$ & $0.08030607$ & $0.01039306$ & $2.7736$ & $0.1294$ & 0.0577 & N \\ 
$1.41451720$ & $0.00145098$ & $0.07008822$ & $0.01036985$ & $0.1645$ & $0.1480$ & 0.0504 & N \\ 
$1.58281137$ & $0.00152920$ & $0.06630589$ & $0.01033905$ & $-1.7477$ & $0.1559$ & 0.0477 & N \\ 
$1.63275628$ & $0.00163583$ & $0.06188241$ & $0.01032218$ & $-1.3461$ & $0.1668$ & 0.0445 & N \\ 
$1.07507415$ & $0.00166747$ & $0.06067859$ & $0.01031709$ & $2.8570$ & $0.1700$ & 0.0436 & N \\ 
$4.60899785$ & $0.00171384$ & $0.05900802$ & $0.01031210$ & $-3.1032$ & $0.1748$ & 0.0424 & N 
\enddata
\end{deluxetable*}

\newpage
\bibliography{references}
\bibliographystyle{aasjournal}

\end{document}